\documentclass[useAMS,usenatbib]{mn2e}

\usepackage{graphicx}
\usepackage{multirow}
\usepackage{epsfig}
\usepackage{amsmath,amssymb, amstext}

\title[Ultraviolet Study of R Arae using Archival \textit{IUE} Data]{Ultraviolet Study of the Active Interacting Binary Star R Arae using Archival \textit{IUE} Data}
\author[P. A. Reed, et al.]{Phillip A. Reed,$^{1,2}$\thanks{E-mail: preed@kutztown.edu} George E. McCluskey, Jr.,$^{2}$ Yoji Kondo,$^{3}$ \newauthor Jorge Sahade,$^{4}$ Edward F. Guinan,$^{5}$ Alvaro Gim\'{e}nez,$^{6}$ Daniel B. Caton,$^{7}$ \newauthor Daniel E. Reichart,$^{8}$ Kevin M. Ivarsen,$^{8}$ and Melissa C. Nysewander$^{8}$\\
$^{1}$Kutztown University, Kutztown, PA 19530, USA\\
$^{2}$Lehigh University, Bethlehem, PA 18015, USA\\
$^{3}$NASA Goddard Space Flight Center, Greenbelt, MD 20771, USA\\
$^{4}$Facultad de Ciencias Astron\'{o}micas, Paseo del Bosque s/n, B1900FWA-La Plata, Argentina\\
$^{5}$Villanova Uiversity, Villanova, PA 19085, USA\\
$^{6}$Centro de Astrobiologia, CSIC/INTA, Carretera de Torrejon a Ajalvir, 28850 Torrejon de Ardoz  (Madrid), Spain\\
$^{7}$Appalachian State University, Boone, NC 28608, USA\\
$^{8}$University of North Carolina at Chapel Hill, Chapel Hill, NC 27599, USA}
\begin{document}



\maketitle

\label{firstpage}

\begin{abstract}
The eclipsing and strongly interacting binary star system R Arae (HD149730) is in a very active and very short-lived stage of its evolution.  R Ara consists of a B9V primary and an unknown secondary.  We have collected the \textit{International Ultraviolet Explorer (IUE)} archival data on R Ara, with most of the data being studied for the first time.  There are 117 high resolution \textit{IUE} spectra taken in 1980, 1982, 1985, 1989, and 1991.  We provide photometric and spectroscopic evidence for mass transfer and propose a geometry for the accretion structure.  We use colour scale radial velocity plots to view the complicated behavior of the blended absorption features and to distinguish the motions of hotter and cooler regions within the system.  We observed a primary eclipse of R Ara in 2008 and have verified that its period is increasing.  A model of the system and its evolutionary status is presented.
\end{abstract}


\section{Introduction}

R Ara is an eclipsing binary with a period of just over 4.4 days and an apparent visual magnitude of 6.7.  A secondary eclipse has never been clearly seen at any wavelength.  Spectroscopically, R Ara is single-lined at best, as the few lines that do follow the orbital motion of the primary deviate from it for much of the orbit.  R Ara's spectrum is plagued with blended absorption features and non-orbital motions.

The \textit{Hipparcos} data on R Ara indicate a parallax of $12.44\pm2.03$ mas, which puts it at a distance of roughly 80 pc.  Its proper motion is reported as $-4.41\pm1.75 \frac{mas}{yr}$ in right ascension and $-13.28\pm1.49 \frac{mas}{yr}$ in declination.  R Ara might be part of the Upper Centaurus-Lupus (UCL) O-B association, but on the southern most edge.  Since the average distance to UCL members is about 140 pc, R Ara would be either on the nearest edge of UCL or in the foreground.  Also, the proper motions of UCL members are somewhat greater than that of R Ara but in the same direction.  If a member of UCL, R Ara's age could be estimated as roughly 15 million years.

R Ara was discovered in 1892 by Alexander W. Roberts at Lovendale, at the time it was called ``Lacaille 6887'', or ``No. (5949) in Chandler's Catalog''.  It was reported as ``(5949) - Arae ... an \textit{Algol}-variable'' by \citet{Roberts1894}. The middle of the $20^{th}$ Century saw more interest in R Ara with photometric studies by \citet{Hertzsprung1942} and \citet{Payne-Gaposchkin1945}, and the first (and extremely useful) spectroscopic analysis by \citet{Sahade1952}.

Hertzsprung found the orbital period of R Ara to be $4^{d}.42507$ with a primary minimum at JD 2,425,818.028 and Payne-Gaposchkin measured a $4^{d}.42509$ period near another primary minimum at JD 2,429,433.348.  Sahade's spectral analysis used 118 plates obtained between 1946 and 1951, and he estimated the semi-amplitude of the primary star to be $K_1\approx60\frac{km}{s}$ and the masses of the components to be $M_1=4M_{\odot}$ and $M_2=1.4M_{\odot}$.  Sahade's analysis determined the B9 classification of the primary.

The \textit{Hipparcos} data provide support for Sahade's classification determination of B9.  A star with R Ara's apparent magnitude, at a distance of 80 pc, would have an absolute magnitude of +2.2, which corresponds to a B9 main-sequence star.  R Ara's colour of B-V = -0.060 is also indicative of a B9 star.  In addition, the ZAMS model of such a B9 main sequence star yields a mass of $3.5M_{\odot}$ to $4M_{\odot}$.

\begin{figure*}
	\includegraphics[width=.85\textwidth]{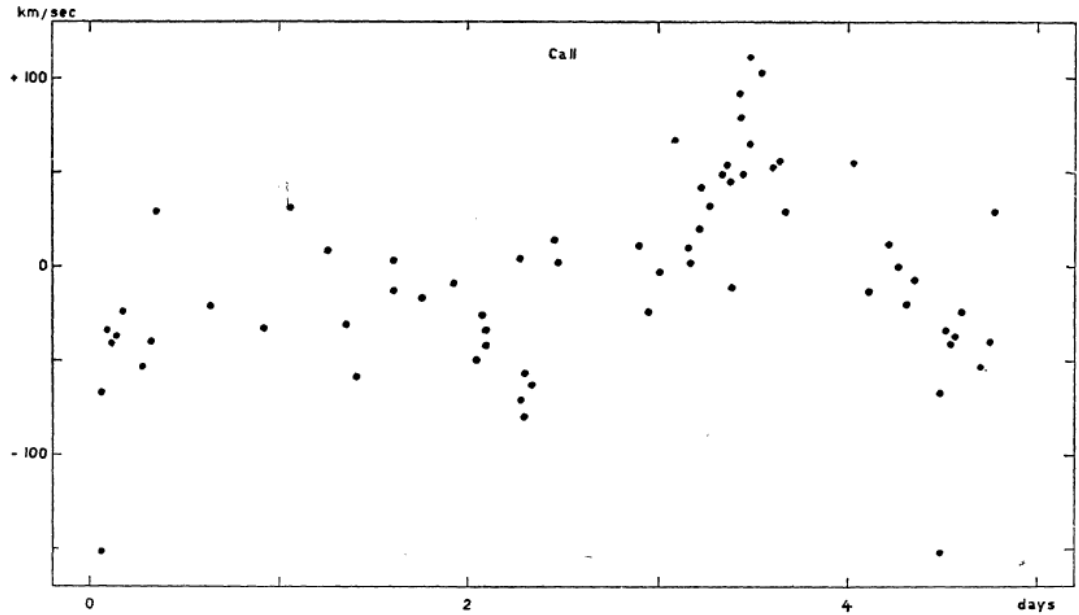}
	\caption{Sahade's RV curve for Ca \textsc{ii}.  This is ``figure 3'' in the paper by \citet{Sahade1952} and was reproduced by permission of the AAS.  The vertical axis has units of $\frac{km}{s}$ with the labeled tickmarks being -100, 0, and +100.  The horizontal axis has units of days and shows labelled tickmarks of 0, 2, and 4.}
	\label{fig:sahadeRV}
\end{figure*}

An original radial velocity (RV) plot by \citet{Sahade1952} is shown in figure \ref{fig:sahadeRV}.  These RVs appear to follow orbital motion of roughly 60 $\frac{km}{s}$ (Sahade's estimate for $K_{1}$) during the last quarter of the cycle, although there is no clear orbital motion seen in the plot.
In Section \ref{sec:UV Spectroscopic Analysis} we provide a $K_{1} = 100\frac{km}{s}$ curve on our colour scale RV plots, not as a fit to the data but as an aid in comparing and contrasting the observed gas motions with estimated orbital motion.

The first \textit{IUE} observations were discussed by \citet{McCluskey1983}, \citet{Kondo1984}, and \citet{Kondo1985}.  They reported P-Cygni line profiles in the 1980 data that exhibit orbital variations, along with the presence of C \textsc{iv} and N \textsc{v} indicating high temperature regions.  Both findings suggest mass transfer processes.  McCluskey and Kondo also reported peculiar light variations seen in the 1982 \textit{IUE} data and in the photometric UV data from the Astronomical Netherlands Satellite \citep{Kondo1981}, and the presence of X-rays as seen with the Einstein Observatory \citep{McCluskey1984}.

The results of these UV analyses spurred more ground-based photometric observations, as reported by groups in New Zealand \citep{Nield1986,Forbes1988,Banks1990,Nield1991} who analysed data taken at Black Birch outstation of Carter Observatory.  Nield determined an ephemeris for the year 1986, for R Ara, of
	\[{Pr. min.} = {JD}\ 2,446,585.1597 + 4.425132E \ ,
\]
and she found that a phase shift for primary minimum may have occurred.  We are very fortunate to have her analysis because her ephemeris fits the \textit{IUE} data better than the earlier (1952) ephemeris by Payne-Gaposchkin, and therefore gives us more accurate orbital phase values than were used in the analysis of McCluskey and Kondo in the early 1980s.

\begin{figure*}
	\includegraphics[width=.85\textwidth]{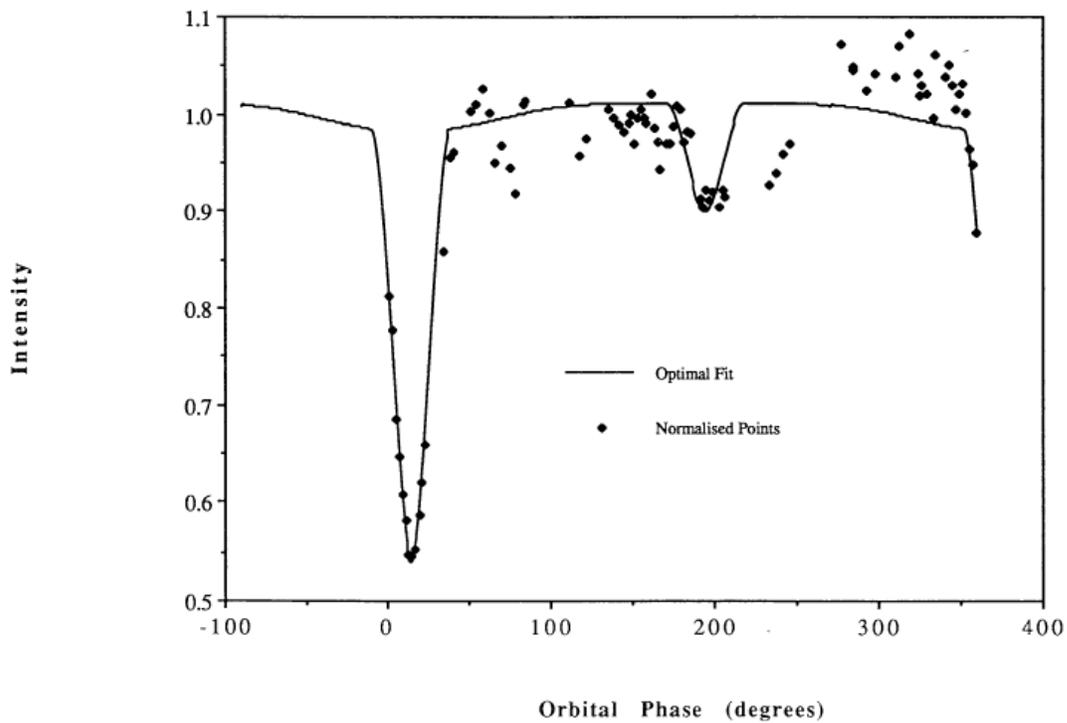}
	\caption{Nield's light curve of visual intensity.  This is ``figure 5'' in the paper by \citet{Nield1991} and was reproduced by permission of Kathy Nield.  The vertical axis is relative intensity and the horizontal axis is the orbital phase (found using the \citet{Payne-Gaposchkin1945} ephemeris) in units of degrees.  The solid line represents the optimal fit of her model.}
	\label{fig:nieldRV}
\end{figure*}

One of Nield's original visual light curves is shown in figure \ref{fig:nieldRV}.  See \citet{Nield1991}, pages 246-247, for the parameters of her model.  While her model provides an excellent fit to the primary eclipse, the outside-of-eclipse portion of the light curve is left essentially un-modeled.  Also, Nield's model does not allow for Roche lobe overflow (RLOF) and therefore does not give R Ara the status of an interacting system at all.  \citet{Banks1990} produced a similar model using visual data with an eccentricity-shifted (e = 0.42) secondary eclipse and a hot spot to explain the increase in brightness before the primary eclipse, but this does not account for the other variations, especially those around first quadrature, nor does it account for RLOF.  It should be noted that such high eccentricities are quite unusual for systems with periods shorter than 5-7 days because these systems are so close that tidal dampening would quickly set the eccentricity to zero.  While it might be possible for extremely rapid mass transfer to drive a non-zero eccentricity, the high eccentricity reported by Banks is most likely in fact spurious.

A recent paper by \citet{Arias1998} describes some optical spectroscopic observations of R Ara taken in 1991.  The description of the data is similar to that of \citet{Sahade1952}, characteristic of badly blended absorption and emission lines.  Arias suggests that the entire system is engulfed in a circumbinary envelope.

This paper focuses on \textit{IUE} data taken in 1985 and 1989, none of which has ever been studied before.  These observations were well planned as they reveal R Ara in the UV during critical parts of its orbit.  In section \ref{Sec:UV Light Curve Analysis} we make sense of R Ara's strange photometric variations by modeling it with a cool accretion structure surrounding the hot primary star.  Section \ref{sec:UV Spectroscopic Analysis} tracks the motions of both hot (too hot for B9 stars) and cool (too cool for B9 stars) regions.  We provide the most detailed evidence to date for the existence of a transient quasi-disc-like accretion structure delivering matter from the less massive secondary star to the more massive primary in R Ara.  We have detected a hot region on the primary that is due to the impact of the mass-transfer stream which drives a ``hot river'' around the primary's equator.

\section{Observations}

The high resolution \textit{IUE} images are listed in Tables \ref{tab:1980IUE} through \ref{tab:1991IUE}, and the orbital phase coverage is also given in those tables.  The 1989 data are particularly useful because there are 36 consecutive spectra spanning one orbital cycle.  Such coverage is not obtainable from the ground, and is free from many secular variations which are prominent for R Ara.  Because these 1989 images are fairly evenly distributed through the orbit, they were used to construct UV light curves and RV plots.  The \textit{IUE} data from 1985 are concentrated near secondary eclipse, another beneficial circumstance.  The short-wave primary (SWP) camera detected wavelengths of 1150 $\sim$ 2000 \AA\ while the long-wave primary (LWP) and long-wave redundant (LWR) cameras covered 1850 $\sim$ 3400 \AA.

All of the \textit{IUE} data are available in the \textit{NEWSIPS} archive.  The \textit{IUE} data used in this paper were obtained from the Multimission Archive at the Space Telescope Science Instute (MAST).  STScI is operated by the Association of Universities for Research in Astronomy, Inc., under NASA contract NAG5-26555. Support for MAST for non-HST data is provided by the NASA Office of Space Science via grant NAG5-7584 and by other grants and contracts.

We also observed a recent primary eclipse of R Ara in 2008 in order to check for an ongoing phase shift and apparent period increase.  The observations were made with the PROMPT5 telescope \citep{Reichart2005}.  Twelve sets of ten, one-second exposures were taken at 30 minute intervals, in the V filter.  The time of minimum light was calculated by the method of \citet{Kwee1956}, using a pregram based on one by \citet{Ghedini1982}.  The resulting time was HJD 2454541.757 $\pm$ 0.002.

\begin{table}
\caption{High resolution \textit{IUE} observations from 1980.}
	\centering
		\begin{tabular}{cccc}
		\hline
		 & \textbf{Observation} & \textbf{Exposure} & \textbf{Orbital}\\
		\textbf{Data ID}&\textbf{Start Time (UT)}&\textbf{Time ($s$)}&\textbf{Phase}\\
		\hline
SWP09026	&	5/17/1980 20:09	&	1080		&	0.074	\\
SWP09138	&	5/27/1980 09:08	&	1800		&	0.231	\\
SWP09144	&	5/27/1980 22:52	&	3000		&	0.362	\\
SWP09151	&	5/29/1980 22:20	&	3300		&	0.809	\\
LWR07781	&	5/17/1980 19:57	&	360	  	&	0.071	\\
LWR07782	&	5/17/1980 20:47	&	660	  	&	0.079	\\
LWR07869	&	5/27/1980 08:42	&	1200		&	0.226	\\
LWR07877	&	5/27/1980 22:08	&	1800		&	0.353	\\
LWR07888	&	5/29/1980 21:42	&	2040		&	0.802	\\
LWR07889	&	5/29/1980 23:20	&	1500		&	0.816	\\
\hline
		\end{tabular}
	\label{tab:1980IUE}
\end{table}
\begin{table}
\caption{High resolution \textit{IUE} observations from 1982.}
	\centering
		\begin{tabular}{cccc}
		\hline
		 & \textbf{Observation} & \textbf{Exposure} & \textbf{Orbital}\\
		\textbf{Data ID}&\textbf{Start Time (UT)}&\textbf{Time ($s$)}&\textbf{Phase}\\
		\hline
SWP17148	&	6/08/1982 07:01	&	3300		&	0.892	\\
SWP17149	&	6/08/1982 08:36	&	2700		&	0.906	\\
SWP17166	&	6/08/1982 18:38	&	3600		&	0.001	\\
SWP17167	&	6/08/1982 20:39	&	4200		&	0.021	\\
SWP17171	&	6/09/1982 07:27	&	2700		&	0.121	\\
SWP17172	&	6/09/1982 08:45	&	2700		&	0.133	\\
SWP17185	&	6/10/1982 18:39	&	2400		&	0.452	\\
SWP17186	&	6/10/1982 20:00	&	2700		&	0.465	\\
SWP17187	&	6/10/1982 21:13	&	2100		&	0.476	\\
SWP17188	&	6/11/1982 07:32	&	2700		&	0.574	\\
SWP17189	&	6/11/1982 08:57	&	3300		&	0.588	\\
SWP17194	&	6/11/1982 18:43	&	2100		&	0.678	\\
SWP17195	&	6/11/1982 19:55	&	1800		&	0.689	\\
SWP17196	&	6/11/1982 20:57	&	2400		&	0.700	\\
LWR13439	&	6/08/1982 06:27	&	1800		&	0.884	\\
LWR13440	&	6/08/1982 08:00	&	1500		&	0.898	\\
LWR13441	&	6/08/1982 09:25	&	1500		&	0.912	\\
LWR13446	&	6/08/1982 19:44	&	3000		&	0.011	\\
LWR13451	&	6/09/1982 06:57	&	1500		&	0.115	\\
LWR13452	&	6/09/1982 08:16	&	1500		&	0.127	\\
LWR13469	&	6/10/1982 19:28	&	1500		&	0.458	\\
LWR13471	&	6/11/1982 07:01	&	1500		&	0.567	\\
LWR13472	&	6/11/1982 08:21	&	1800		&	0.580	\\
LWR13477	&	6/11/1982 19:24	&	1500		&	0.684	\\
LWR13478	&	6/11/1982 20:29	&	1500		&	0.694	\\
\hline
		\end{tabular}
	\label{tab:1982IUE}
\end{table}
\begin{table}
\caption{High resolution \textit{IUE} observations from 1985.}
	\centering
		\begin{tabular}{cccc}
		\hline
		 & \textbf{Observation} & \textbf{Exposure} & \textbf{Orbital}\\
		\textbf{Data ID}&\textbf{Start Time (UT)}&\textbf{Time ($s$)}&\textbf{Phase}\\
		\hline
SWP26253	&	6/25/1985 05:32	&	2400		&	0.394	\\
SWP26255	&	6/25/1985 08:10	&	2520		&	0.419	\\
SWP26256	&	6/25/1985 09:31	&	2520		&	0.432	\\
SWP26257	&	6/25/1985 10:48	&	2580		&	0.444	\\
SWP26258	&	6/25/1985 12:08	&	2700		&	0.457	\\
SWP26259	&	6/25/1985 13:29	&	3000		&	0.470	\\
SWP26260	&	6/25/1985 15:37	&	2700		&	0.490	\\
SWP26261	&	6/25/1985 16:57	&	2700		&	0.502	\\
SWP26262	&	6/25/1985 18:25	&	2640		&	0.516	\\
SWP26263	&	6/25/1985 19:43	&	2700		&	0.528	\\
SWP26315	&	6/30/1985 06:06	&	3000		&	0.530	\\
SWP26316	&	6/30/1985 07:39	&	3360		&	0.545	\\
SWP26317	&	6/30/1985 09:16	&	3300		&	0.561	\\
SWP26318	&	6/30/1985 10:50	&	2700		&	0.575	\\
SWP26319	&	6/30/1985 12:11	&	2400		&	0.587	\\
SWP26324	&	6/30/1985 19:07	&	2100		&	0.652	\\
SWP26325	&	6/30/1985 20:15	&	1920		&	0.662	\\
SWP26643	&	9/06/1985 19:53	&	3600		&	0.028	\\
SWP26645	&	9/06/1985 22:57	&	2400		&	0.055	\\
LWP06267	&	6/25/1985 06:18	&	1500		&	0.400	\\
LWP06269	&	6/25/1985 08:58	&	1380		&	0.425	\\
LWP06270	&	6/25/1985 10:19	&	1380		&	0.438	\\
LWP06271	&	6/25/1985 11:37	&	1440		&	0.450	\\
LWP06272	&	6/25/1985 12:59	&	1500		&	0.463	\\
LWP06273	&	6/25/1985 15:07	&	1500		&	0.483	\\
LWP06274	&	6/25/1985 16:28	&	1500		&	0.496	\\
LWP06275	&	6/25/1985 17:49	&	1440		&	0.509	\\
LWP06276	&	6/25/1985 19:15	&	1440		&	0.522	\\
LWP06277	&	6/25/1985 20:32	&	1020		&	0.534	\\
LWP06301	&	6/30/1985 05:35	&	1500		&	0.524	\\
LWP06302	&	6/30/1985 07:05	&	1680		&	0.538	\\
LWP06303	&	6/30/1985 08:41	&	1680		&	0.553	\\
LWP06304	&	6/30/1985 10:18	&	1500		&	0.568	\\
LWP06305	&	6/30/1985 11:41	&	1380		&	0.581	\\
LWP06306	&	6/30/1985 12:57	&	900	  	&	0.592	\\
LWP06311	&	6/30/1985 19:47	&	1440		&	0.657	\\
LWP06698	&	9/06/1985 21:00	&	1920		&	0.036	\\
\hline
		\end{tabular}
	\label{tab:1985IUE}
\end{table}
\begin{table}
\caption{High resolution \textit{IUE} observations from 1989.}
	\centering
		\begin{tabular}{cccc}
		\hline
		 & \textbf{Observation} & \textbf{Exposure} & \textbf{Orbital}\\
		\textbf{Data ID}&\textbf{Start Time (UT)}&\textbf{Time ($s$)}&\textbf{Phase}\\
		\hline
SWP36973	&	9/10/1989 07:55	&	2400		&	0.977	\\
SWP36974	&	9/10/1989 09:17	&	3600		&	0.991	\\
SWP36975	&	9/10/1989 11:21	&	3000		&	0.010	\\
SWP36976	&	9/10/1989 12:58	&	3600		&	0.026	\\
SWP36977	&	9/10/1989 14:54	&	3600		&	0.044	\\
SWP36978	&	9/10/1989 16:45	&	3300		&	0.061	\\
SWP36986	&	9/11/1989 05:18	&	3000		&	0.179	\\
SWP36987	&	9/11/1989 06:56	&	480	  	&	0.191	\\
SWP36988	&	9/11/1989 07:35	&	3600		&	0.201	\\
SWP36993	&	9/11/1989 15:59	&	3300		&	0.280	\\
SWP36999	&	9/12/1989 04:09	&	3600	  &	0.395	\\
SWP37002	&	9/12/1989 11:22	&	3000		&	0.462	\\
SWP37005	&	9/12/1989 18:55	&	3600		&	0.534	\\
SWP37006	&	9/12/1989 20:42	&	3300		&	0.551	\\
SWP37013	&	9/13/1989 10:26	&	3000		&	0.679	\\
SWP37017	&	9/13/1989 17:36	&	3300		&	0.747	\\
SWP37020	&	9/14/1989 00:07	&	3600		&	0.809	\\
SWP37023	&	9/14/1989 07:03	&	3600		&	0.874	\\
SWP37025	&	9/14/1989 12:43	&	3000		&	0.927	\\
LWP16314	&	9/10/1989 08:44	&	1200  	&	0.983	\\
LWP16315	&	9/10/1989 10:25	&	2400		&	0.001	\\
LWP16316	&	9/10/1989 12:19	&	1800		&	0.018	\\
LWP16317	&	9/10/1989 14:08	&	2100		&	0.035	\\
LWP16318	&	9/10/1989 16:03	&	1920		&	0.053	\\
LWP16319	&	9/10/1989 17:49	&	1800		&	0.069	\\
LWP16326	&	9/11/1989 06:16	&	1800		&	0.187	\\
LWP16330	&	9/11/1989 15:22	&	1800		&	0.272	\\
LWP16336	&	9/12/1989 05:22	&	1800		&	0.404	\\
LWP16339	&	9/12/1989 12:22	&	1500		&	0.470	\\
LWP16342	&	9/12/1989 18:15	&	1920		&	0.526	\\
LWP16343	&	9/12/1989 20:03	&	1800		&	0.542	\\
LWP16349	&	9/13/1989 11:25	&	1800		&	0.687	\\
LWP16353	&	9/13/1989 18:40	&	1800		&	0.755	\\
LWP16355	&	9/14/1989 01:15	&	1800		&	0.817	\\
LWP16358	&	9/14/1989 08:12	&	1800		&	0.883	\\
LWP16360	&	9/14/1989 13:42	&	1800		&	0.935	\\
\hline
		\end{tabular}
	\label{tab:1989IUE}
\end{table}
\begin{table}
	\caption{High resolution \textit{IUE} observations from 1991.}
	\centering
		\begin{tabular}{cccc}
		\hline
		 & \textbf{Observation} & \textbf{Exposure} & \textbf{Orbital}\\
		\textbf{Data ID}&\textbf{Start Time (UT)}&\textbf{Time ($s$)}&\textbf{Phase}\\
		\hline
SWP41934	&	6/27/1991 14:08	&	1800		&	0.053	\\
SWP41936	&	6/27/1991 18:35	&	3000		&	0.096	\\
SWP41937	&	6/27/1991 20:05	&	1920		&	0.109	\\
SWP41955	&	6/29/1991 13:30	&	1380		&	0.498	\\
SWP41958	&	6/29/1991 18:00	&	2700		&	0.543	\\
LWP20701	&	6/27/1991 14:48	&	900		  &	0.058	\\
LWP20703	&	6/27/1991 19:35	&	1320		&	0.104	\\
LWP20713	&	6/29/1991 17:20	&	900	  	&	0.534	\\
\hline
		\end{tabular}
\label{tab:1991IUE}
\end{table}

\section{UV Photometric Analysis}
\label{Sec:UV Light Curve Analysis}

\begin{figure*}
	\includegraphics[angle=90, width=.90\textwidth]{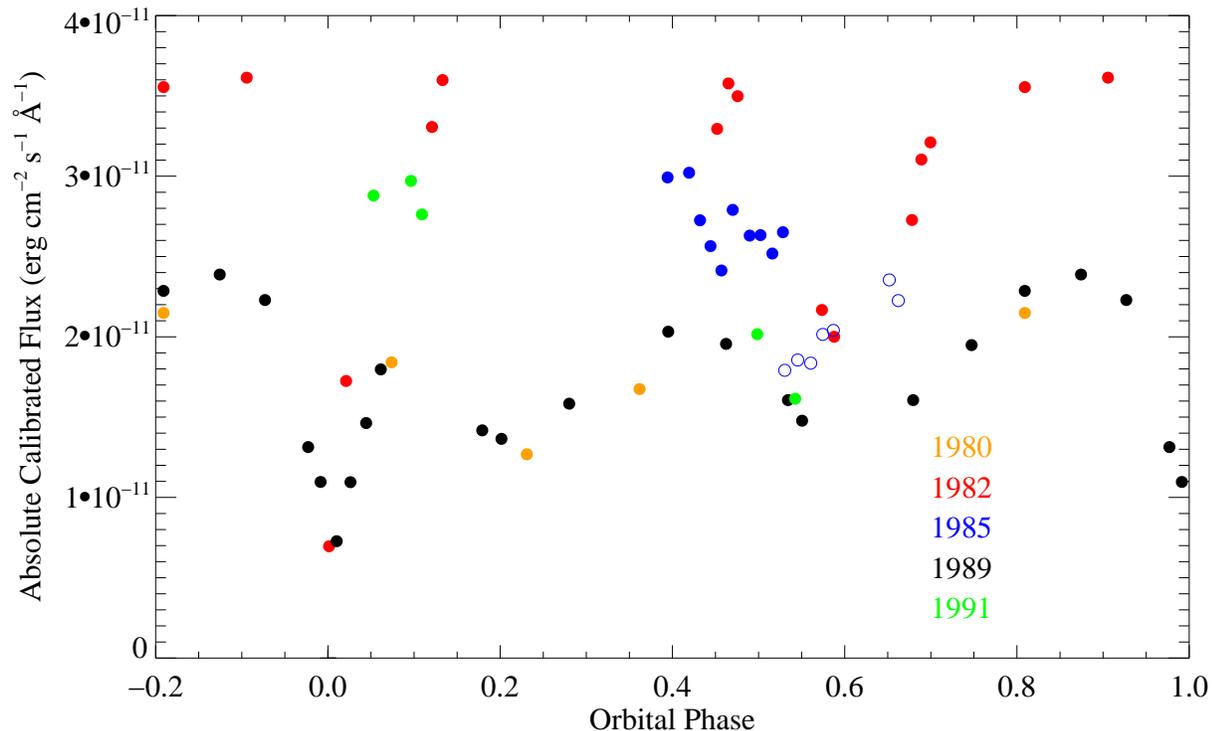}
	\caption{The \textit{IUE} light curves of absolute calibrated flux at 1320 \AA.  The 1985 data were taken during two consecutive orbits, rather than consectuive images during a single orbit.  The filled blue circles are data taken on 25 June 1985 and the open blue circles are data taken on 30 June 1985, one cycle later.}
	\label{fig:LCall}
\end{figure*}

In our photometric analysis of the \textit{IUE} data, continuum levels were found for ``flat'' regions of the spectrum that are free of spectral features.  These local continua were used to produce our light curves.  The local continuum levels near 1320 \AA\ for all of the \textit{IUE} data are plotted in figure \ref{fig:LCall}.  These are un-normalised light curves of absolute calibrated flux.

R Ara exhibits significant photometric variations on different time scales.  The secular variations are quite evident in the  comparison between 1982 and 1989, for example, as the primary minimum reaches a similar value while the system as a whole is much brighter in 1982 than in 1989.   The 1985 data show that these variations can also take place from one orbit to the very next.  These variations confuse light curves from ground-based images because they must be folded together with data spanning several orbits.  Observing from space can avoid this by enabling consecutive images from a single orbit.

The 1989 \textit{IUE} data, being studied for the first time in this paper, are extremely useful because they offer good phase coverage of an entire 4.4 day period with consecutive images. These data are free of secular variations.

\begin{figure}
	\includegraphics[angle=90,width=.475\textwidth]{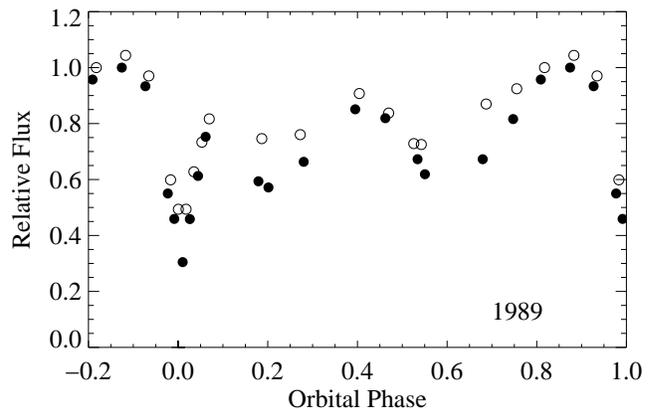}
	\caption{The 1989 \textit{IUE} light curve of the continuum fluxes measured at 1320 \AA\ (solid circles) and 2915 \AA\ (open circles), and normalised to 1.0 at orbital phases 0.809 and 0.817, respectively.}
	\label{fig:1989LC}
\end{figure}

\begin{figure}
	\includegraphics[angle=90,width=.475\textwidth]{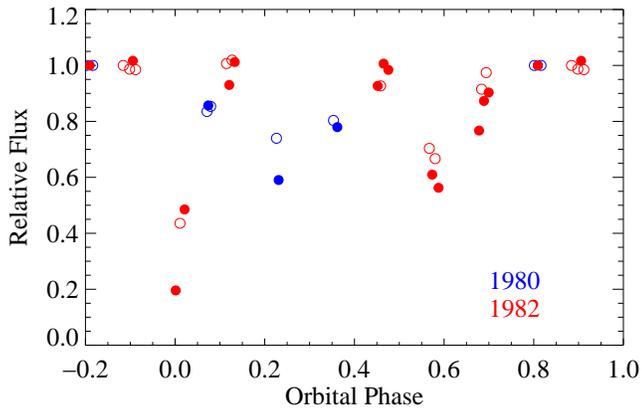}
	\caption{The 1980 and 1982 \textit{IUE} light curves, normalised as in figure \ref{fig:1989LC}.  The filled circles are relative fluxes at 1320 \AA, and the open circles are at 2915 \AA.}
	\label{fig:1980-82LC}
\end{figure}

Our light curves of the 1989 data at 1320 \AA\ and 2915 \AA\ are shown in figure \ref{fig:1989LC}.
In addition to the primary eclipse, there are two other ``dips'' in the light curve.  The 1980 and 1982 data (normalized and plotted together in figure \ref{fig:1980-82LC}) also show both dips.  The 1980 observations cover the first dip and the second dip was observed  in 1982, but together they are consistent with the single orbit of 1989.  Is one of these two dips the secondary eclipse?  If so, which one? 

It was noted by \citet{Kondo1985} that the second of these dips (near phase 0.6) becomes shallower with increasing wavelength.  Our new 1989 light curves (figure \ref{fig:1989LC}) confirm this and extend the condition to the first dip as well.  This means that neither of the dips can be the result of the secondary eclipse, and both dips are caused by something else; they are caused by something cooler eclipsing the primary and not the primary eclipsing the secondary.  We suspect, of course, that a cool accretion structure surrounding the B9 primary is responsible for these dips in the light curve.  What is the geometry of this accretion structure that causes the variations in the light curve?

In the attempt to model cool regions (or clouds) located in the line of sight to the primary star, we chose to use the ``spots'' feature in the binary star modeling program {\small{BINARY MAKER 3}} (BM3) by \citet{Bradstreet2002}.  We started with the parameters given by \citet{Nield1991}, which as previously noted were a good fit for the primary eclipse but not for the rest of the light curve.  Nonetheless, their model of R Ara's primary eclipse is an excellent starting point for understanding its light curve.

\begin{figure}
	\includegraphics[angle=90,width=.475\textwidth]{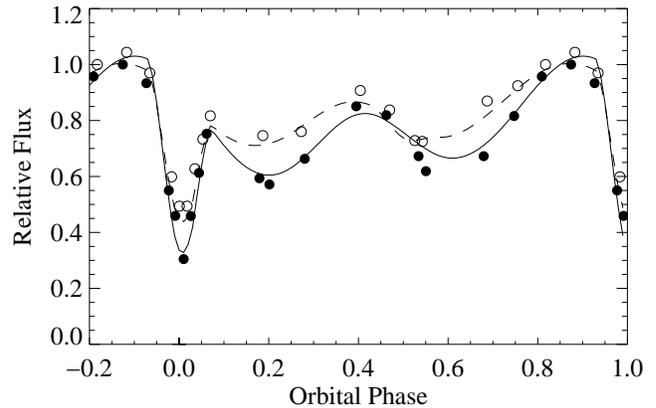}
	\caption{The 1989 \textit{IUE} light curves with our BM3 model.  The solid circles are data points at 1320 \AA\ and the open circles are at 2915 \AA.  The solid line is the BM3 model at 1320 \AA\ and the dashed line is the model at 2915 \AA.}
	\label{fig:1989-w-model}
\end{figure}

For our model, we first increased the size of the secondary to fill its Roche lobe, and to therefore allow for RLOF.  This, in turn, led to a slightly greater inclination angle.  We kept Nield's parameters for gravity darkening, limb darkening, and reflection effects.  Table \ref{tab:SynModelParams} lists the other parameters used in our model.  Figure \ref{fig:1989-w-model} shows our synthetic light curves plotted over the 1989 \textit{IUE} data, for each end of the UV spectrum, 1320 \AA\ and 2915 \AA.
\begin{table}
	\centering
	\caption{The parameters for our ``best-fit'' BM3 model for R Ara.  The relative radius is the size of the star relative to the distance between the centres of the stars' masses.  The temperature factor for a spot (cloud) is its temperature relative to the surface temperature of the star.}
		\begin{tabular}{|p{2cm}c|c|}
		\hline
	Mass Ratio & q	&	0.39 ($\pm$ 0.05)		\\
	\hline
	Relative Radii&	r$_1$	&	0.192 ($\pm$ 0.025)		\\
	&	r$_2$	&	-1.0 (fill Roche Lobe)		\\
	\hline
	Ratio of the Radii & $\frac{r_1}{r_2}$ & 0.64 ($\pm$ 0.05)  \\
	\hline
	Fractional Luminosities  & L$_1$ & 0.996 ($\pm\ 1.5 \times 10^{-3}$)\\
	(at 1320 \AA) & L$_2$ & 0.004 ($\pm\ 1.5 \times 10^{-3}$)\\
	\hline	
	Fractional Luminosities  & L$_1$ & 0.874 ($\pm\ 1.5 \times 10^{-3}$)\\
	(at 2915 \AA) & L$_2$ & 0.126 ($\pm\ 1.5 \times 10^{-3}$)\\
	\hline
	Surface	&	T$_1$	&	12500 ($\pm$ 2000) K		\\
	Temperatures	&	T$_2$	&	7000 ($\pm$ 1300)	K	\\
	\hline
	Orientation	&	Inclination	&	$78^\circ$ ($\pm$ 4)		\\
	&	Norm. Phase	&	0.85  \\
	\hline
	Spot 1	&	colatitude	&	$80.2^\circ$ ($\pm$ 6)		\\
	&	longitude	&	$139.5^\circ$ ($\pm$ 3.5)		\\
	&	radius	&	$35.7^\circ$ ($\pm$ 1)		\\
	&	temp. factor	&	0.28 ($\pm$ 0.05)	\\
	\hline
	Spot 2	&	colatitude	&	$94.7^\circ$ ($\pm$ 6)		\\
	&	longitude	&	$287.0^\circ$ ($\pm$ 3.5)		\\
	&	radius	&	$40.0^\circ$ ($\pm$ 1.5)		\\
	&	temp. factor	&	0.30 ($\pm$ 0.05)		\\
	\hline
	\end{tabular}
	\label{tab:SynModelParams}
\end{table}

A hot spot, similar to that suggested by \citet{Banks1990} to explain the ``hump'' preceding primary minumum, has effects that are not consistent with the data throughout the UV spectrum and it does not account for the other variations throughout the orbit.

\begin{figure}
	\begin{center}$
	\begin{array}{ccc}	
	\multicolumn{3}{c}{\includegraphics[width=.2\textwidth]{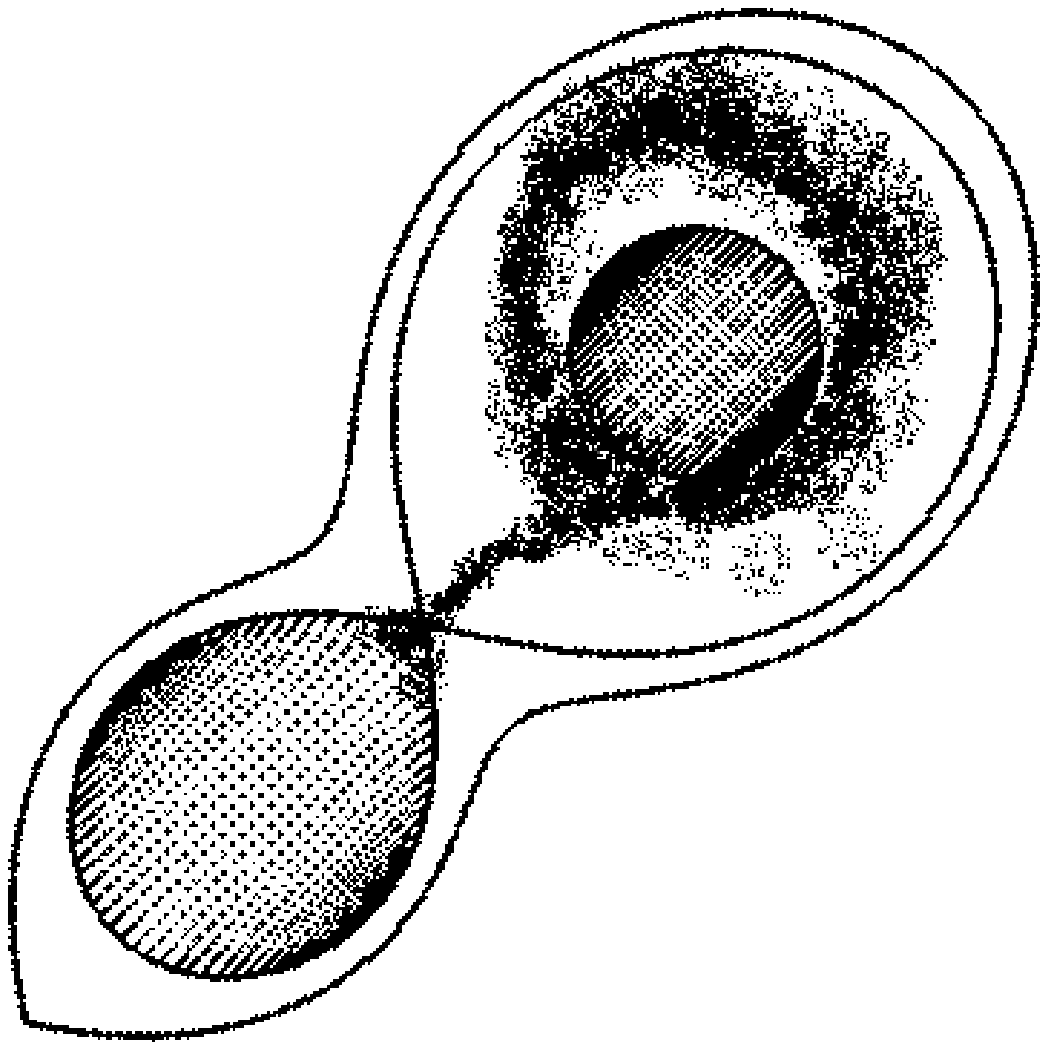} \vspace{0.15in}}\\
	0.000
	&\includegraphics[width=.06\textwidth]{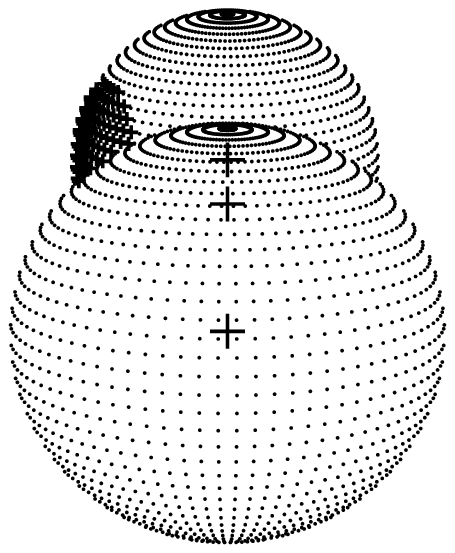}
	\hspace{0.2in}	
	&\includegraphics[width=.09\textwidth]{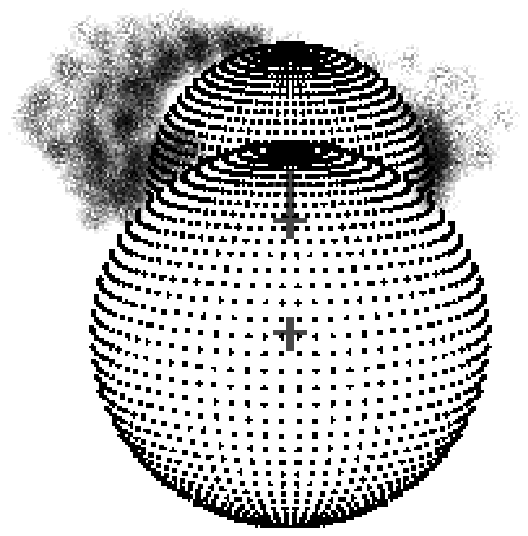}\\
	0.125
	&\includegraphics[width=.13\textwidth]{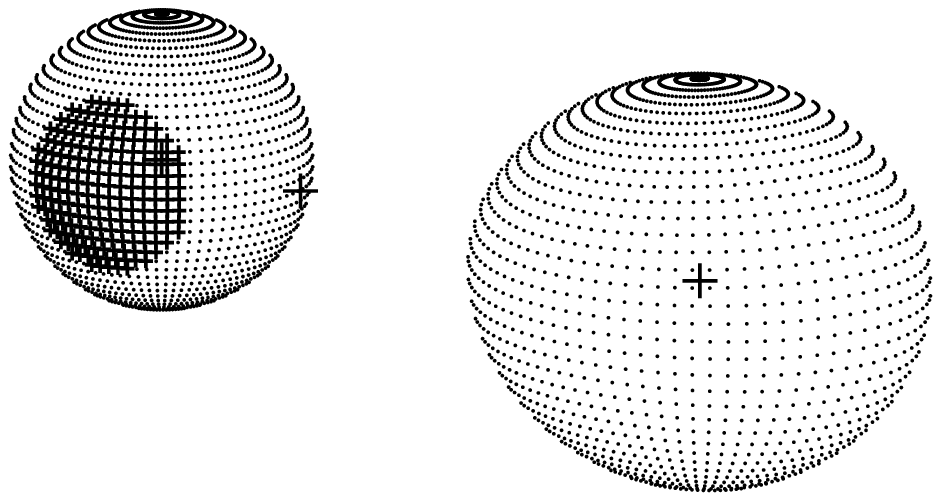}
	\hspace{0.2in}	
	&\includegraphics[width=.16\textwidth]{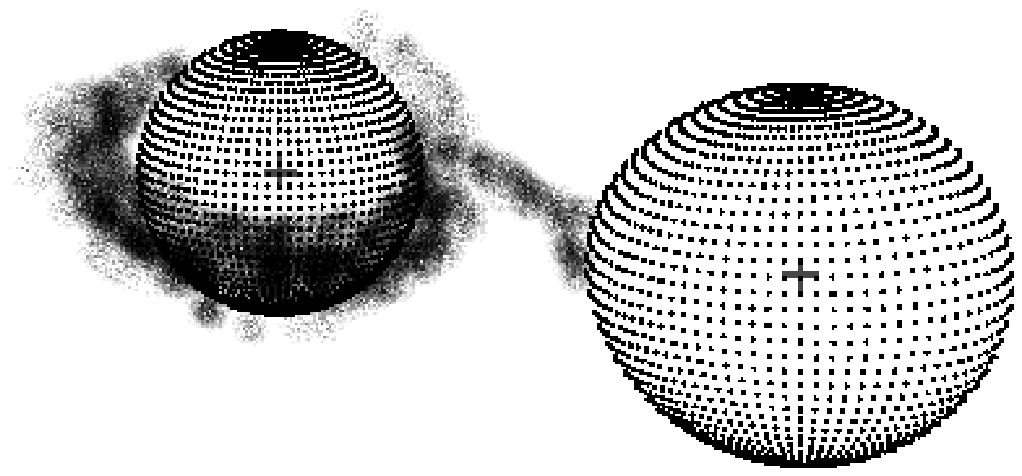}\\
	0.250
	&\includegraphics[width=.16\textwidth]{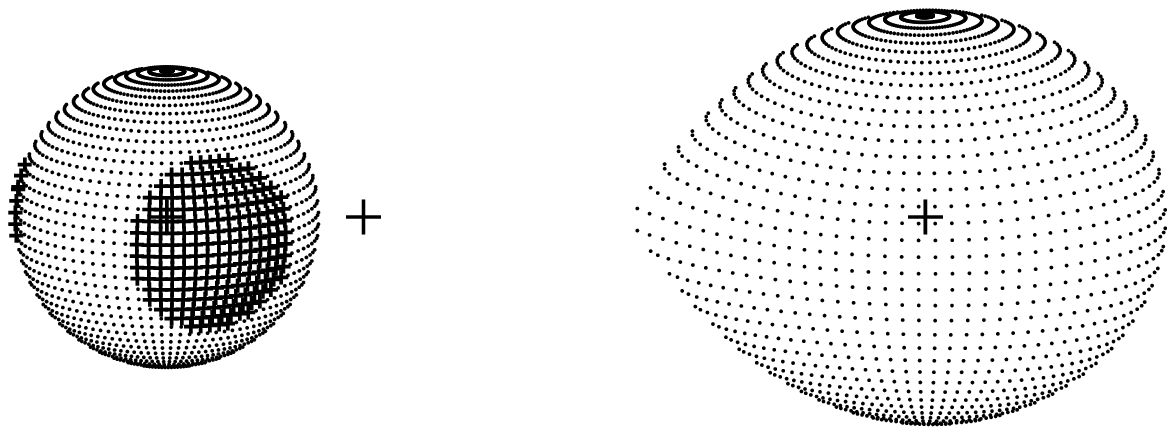}
	\hspace{0.2in}	
	&\includegraphics[width=.19\textwidth]{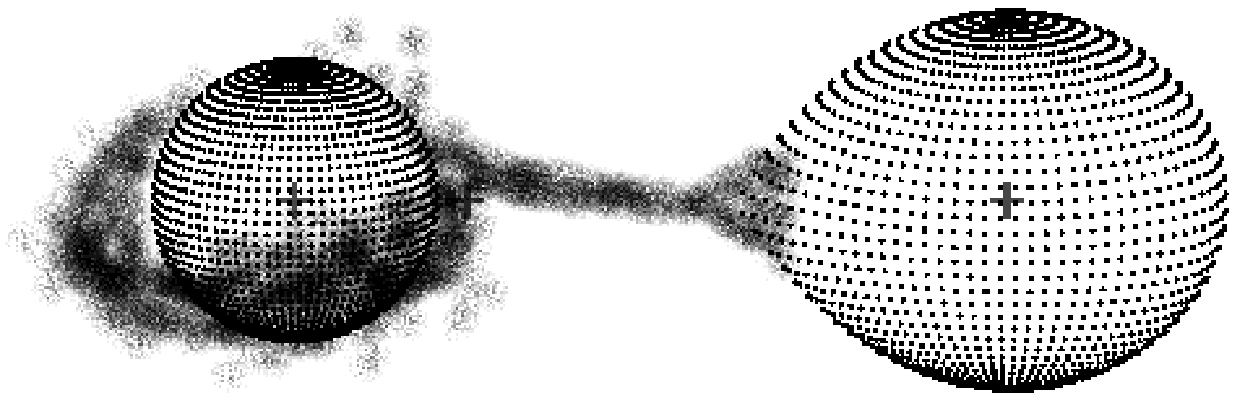}\\
	0.375
	&\includegraphics[width=.13\textwidth]{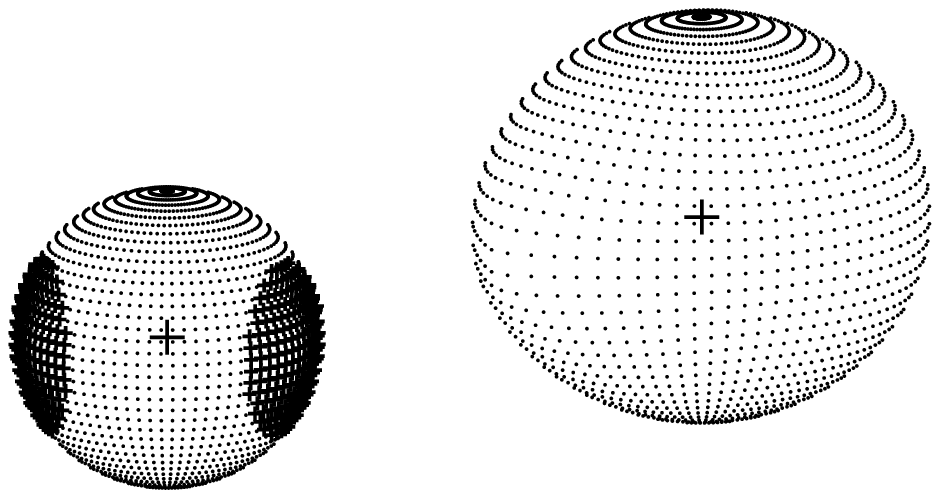}
	\hspace{0.2in}	
	&\includegraphics[width=.17\textwidth]{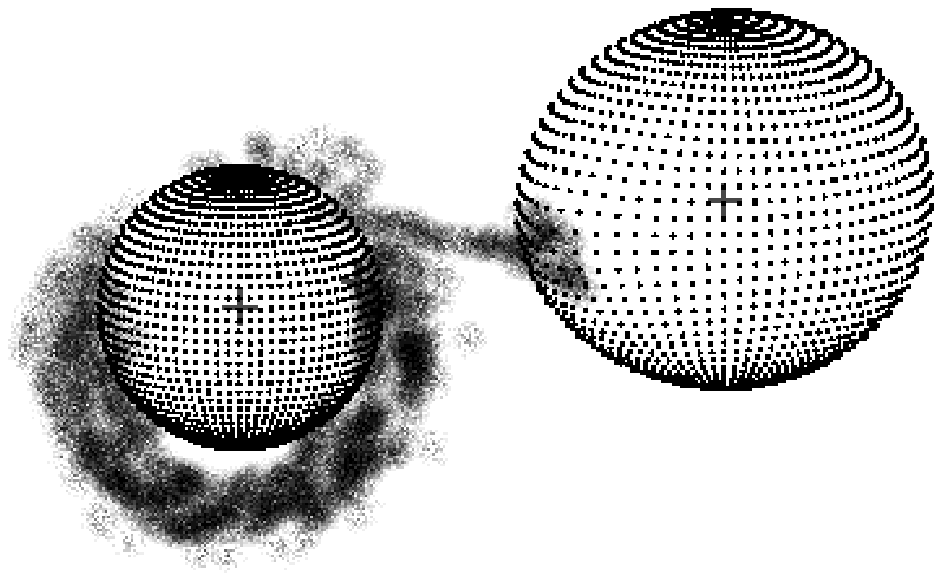}\\
	0.500
	&\includegraphics[width=.06\textwidth]{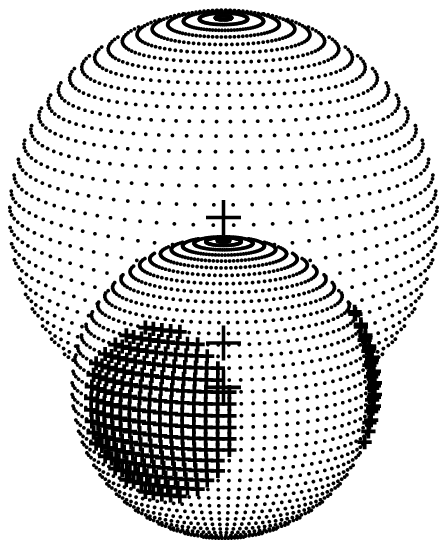}
	\hspace{0.2in}	
	&\includegraphics[width=.10\textwidth]{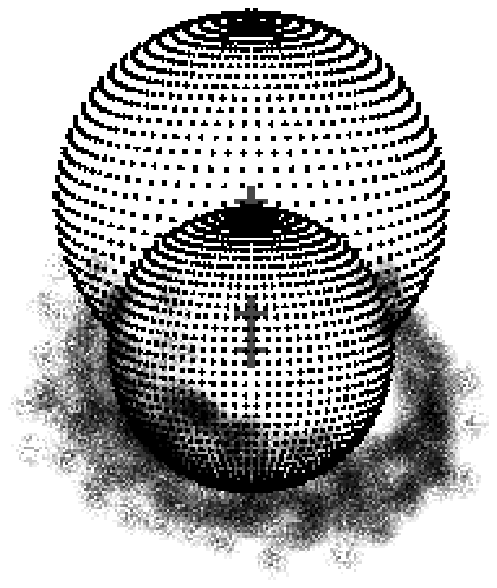}\\	
	\end{array}$
	\end{center}	
	\caption{(TOP) An overhead view illustrating the geometry of the proposed accretion structure.  (BOTTOM) The line-of-sight views for the first half of the orbit, showing (left) the cool regions from the BM3 model and (right) illustrations of the proposed accretion structure.  The phase value for each pair of images is listed to the left.}
	\label{fig:structure}
\end{figure}

Our model's ``cool spots'' should be interpreted as cool clouds in the line of sight to the primary star.  Because BM3 is limited to circular regions located at the photosphere of the star, the clouds of our synthetic model have the same restrictions; however, the model does yield a general geometry and temperature for the (accretion) clouds.  Figure \ref{fig:structure} shows the correlation between the BM3 model and the entire accretion structure.
The cool accretion structure can not be a solid, symmetric ``disc'' or ``annulus'', as such a structure would not cause the phase-dependent variation we see.  It can not be a constant thickness through time either, or we would not see the secular variations that we do.  Rather, R Ara's accretion structure must be more transient and asymmetric.  

The asymmetry of the accretion structure around the primary, proposed in our model (top of figure \ref{fig:structure}), is similar to that found by \citet{Boyarchuk2002} in their hydrodynamic simulations of the interacting system $\beta$ Lyr.  Coupling this asymmetry with the system's inclination may explain the orbital variations in R Ara's light curve.  The cool clouds in the restricted BM3 model would correspond to where the accretion structure is closest to the primary and is therefore in the line of sight to it, while the farther regions of the cloud descend below our line of sight to the star revealing more of its hot equator.  The alternative explanation that would not rely on the inclination and asymmetry would be that there are higher density concentrations in the accretion structure at the location of the ``spots'' in our BM3 model.

The transient nature of the accretion structure, proposed by our model to account for secular variations,  is consistent with the findings of \citet{Richards1998}.  They found that Algol-type binaries with periods shorter than R Ara's, such as $\beta$ Per (P = 2.9 days), exhibit mass-transfer streams that strike the accreting star more directly and do not form stable accretion discs.  At the same time, they find that the mass transfer in Algols with periods longer than R Ara's, like TT Hya (P = 6.9 days), does not create a direct impact and therefore quickly produces a stable accretion disc.  An interacting system like R Ara, with a 4.4 day period, might find itself varying between a ``$\beta$ Per-like'' state and a ``TT Hya-like'' state.

We propose that the accretion rate onto the primary in R Ara is variable.  During times of rapid accretion, the transfer stream strikes the primary directly and produces a thinner accretion structure.  This should correspond to brighter epochs, like 1982.  As the accretion structure builds up, during times of slower accretion onto the primary, it will ``push'' the mass-transfer stream away from the star causing a less direct impact and a thicker accretion structure.  This latter case will cause the system to appear dimmer overall, like in 1989.  As the accretion rate increases, the stream may return to the more direct impact state and the system will become brighter again.

\begin{figure}
	\includegraphics[angle=90,width=.475\textwidth]{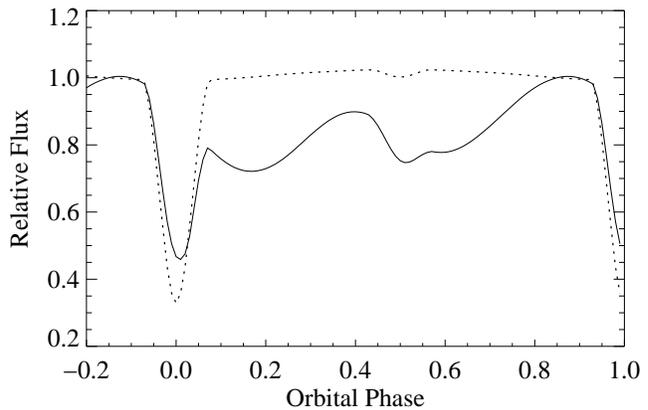}
	\caption{This is our synthetic BM3 light curve under two further conditions.  The solid line is our model with cool clouds eclipsing the primary, as it would look in the visual (5500 \AA).  The dotted line is our model in the ultraviolet (1320 \AA), but with no cool clouds.  For the dotted line, we disabled only the ``spots'' feature in BM3, leaving all other parameters the same.}
	\label{fig:VIS-NS-LC}
\end{figure}

For completeness, we have included synthetic light curves of our BM3 model under two further conditions, in figure \ref{fig:VIS-NS-LC}.  The dotted line is our model, at 1320 \AA, but without any cool clouds (or ``spots'' in BM3) eclipsing the primary.  This is essentially what Nield's model would look like in the UV.  Note that the secondary eclipse should almost vanish at 1320 \AA.  The \textit{IUE} data tell us that secondary ``dips'' in the light curve are in fact deeper in UV.

The solid line in figure \ref{fig:VIS-NS-LC} is our BM3 model, cool clouds included, as viewed at 5500 \AA, in the visual.  Indeed, our model can be extended into the visual region to explain the ``hump'' before primary minumum, the wide and apparently shifted ``secondary eclipse'', and the other variations near first quadrature.

Our observation of the 16 March 2008 primary eclipse of R Ara confirms the ongoing period increase detected by Nield.  In section \ref{sec:Descussion} we estimate R Ara's rates of period change and mass transfer.

We have proposed a model for R Ara to explain its strange variations in brightness.  In the following section we put our photometric model to the test through our analysis of R Ara's spectral features.  Are the motions of the cooler ions consistent with our proposed accretion structure and are the motions of the hotter ions consistent with our proposed ``hot river'' driven around the primary's equator by the impact of the mass-transfer stream?

\section{UV Spectroscopic Analysis}
\label{sec:UV Spectroscopic Analysis}

In section \ref{Sec:UV Light Curve Analysis} we proposed a model for R Ara consisting of an undetected F to K type secondary undergoing RLOF and donating mass to the more massive B9 primary.  The mass-transfer stream strikes the primary, either directly or at more of a glancing angle, which drives a ``hot river'' around the primary's equator.  The imact site and equator are where we would expect to find the hotter ions, like N \textsc{v} and C \textsc{iv}, which require much hotter ($>$ 30,000 K) plasmas than are typically found in late B-type stars.  The ions that are too cool for a B9, such as Ni \textsc{ii} and Fe \textsc{ii}, should be found in the accretion structure.

\begin{figure}
	\begin{center}$
	\begin{array}{ccc}
	A	
	\includegraphics[width=.135\textwidth]{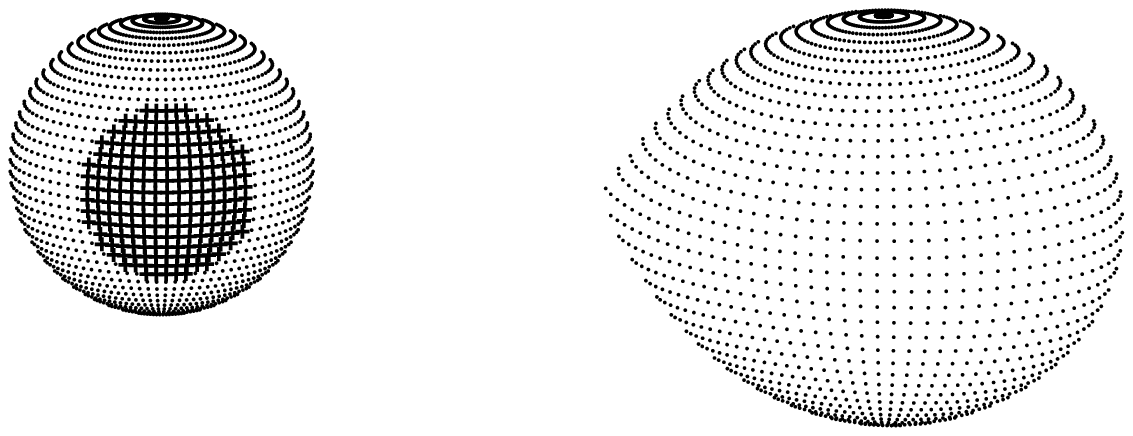}	
	\hspace{0.05in}
	\includegraphics[width=.14\textwidth]{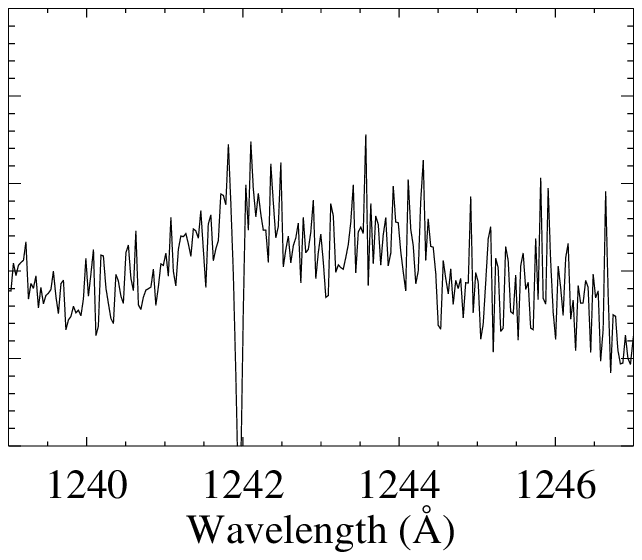}	
	\includegraphics[width=.16\textwidth]{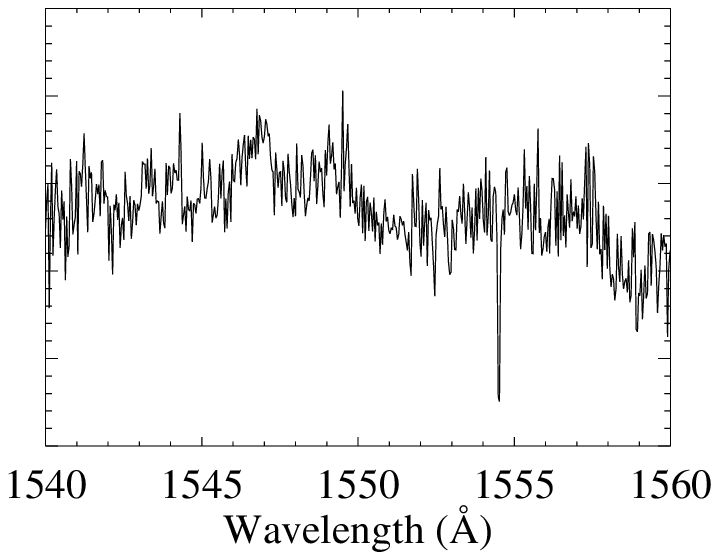}\\
	\hspace{0.10in}
	B
	\includegraphics[width=.110\textwidth]{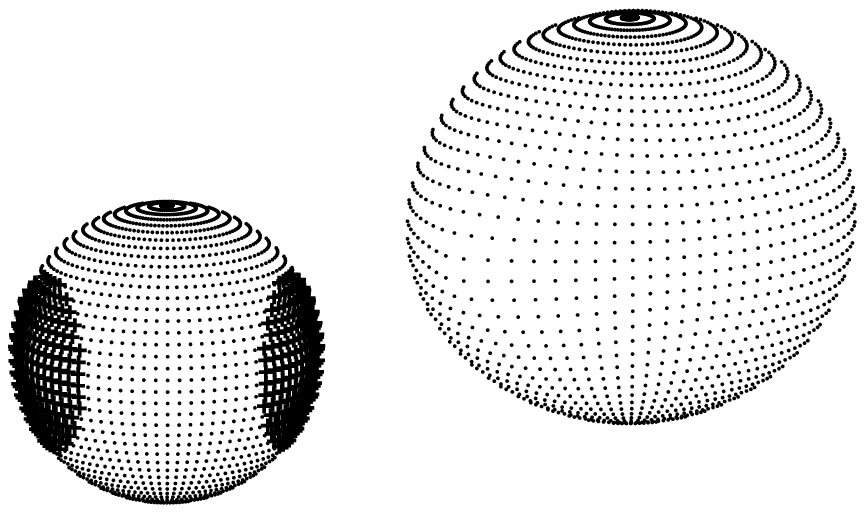}	
	\hspace{0.12in}
	\includegraphics[width=.14\textwidth]{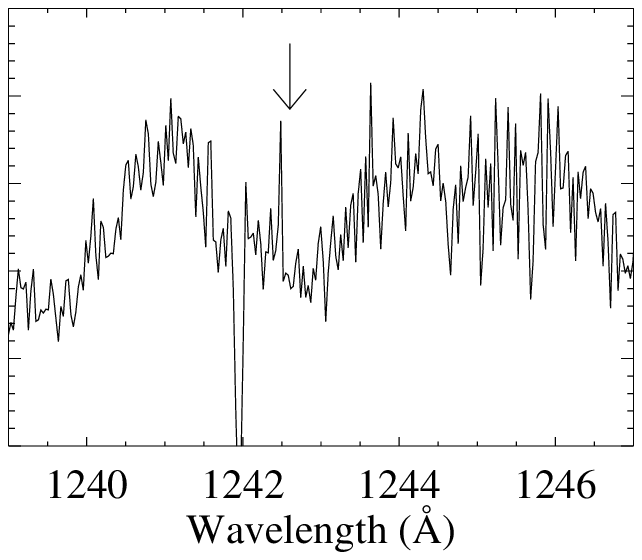}	
	\includegraphics[width=.16\textwidth]{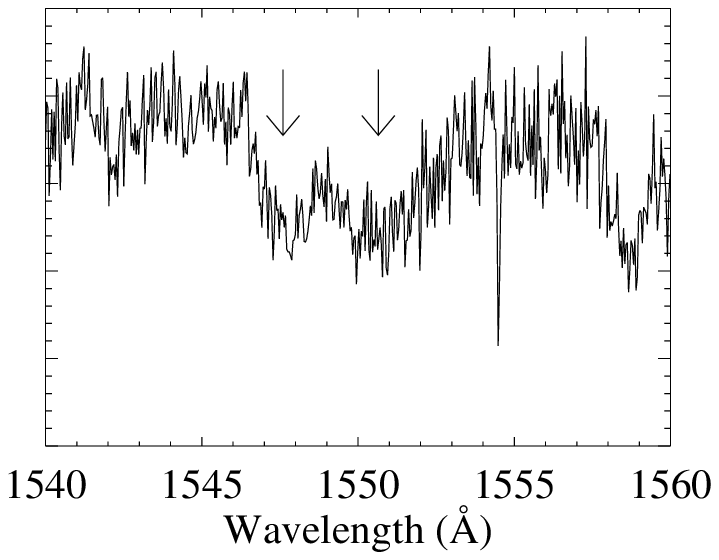}\\
	\hspace{0.17in}
	C
	\includegraphics[width=.070\textwidth]{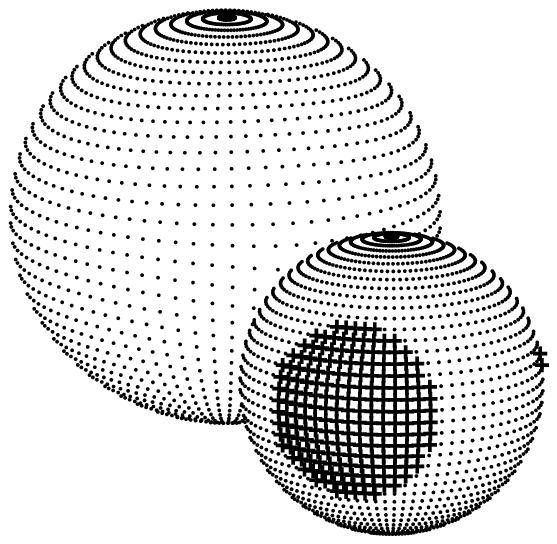}	
	\hspace{0.32in}
	\includegraphics[width=.14\textwidth]{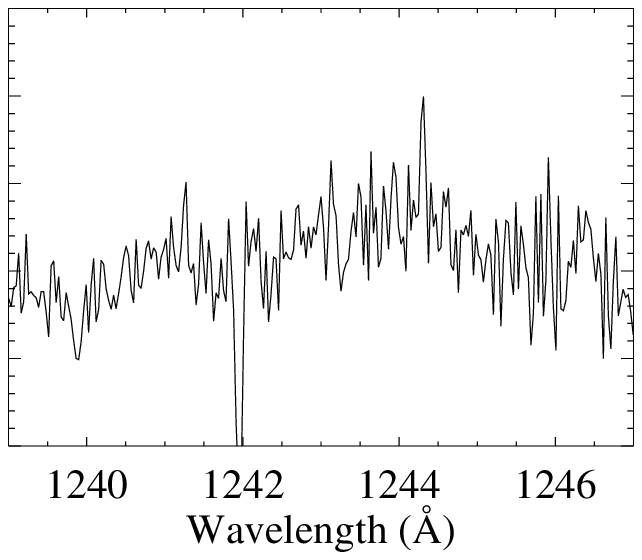}	
	\includegraphics[width=.16\textwidth]{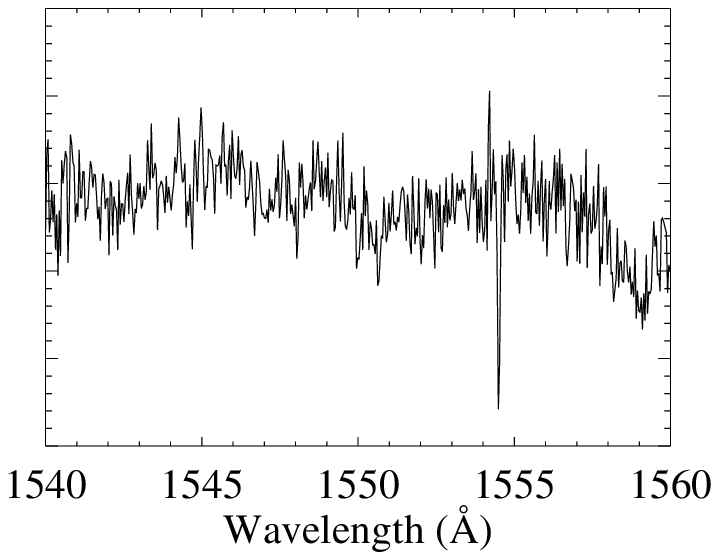}\\
	D
	\includegraphics[width=.135\textwidth]{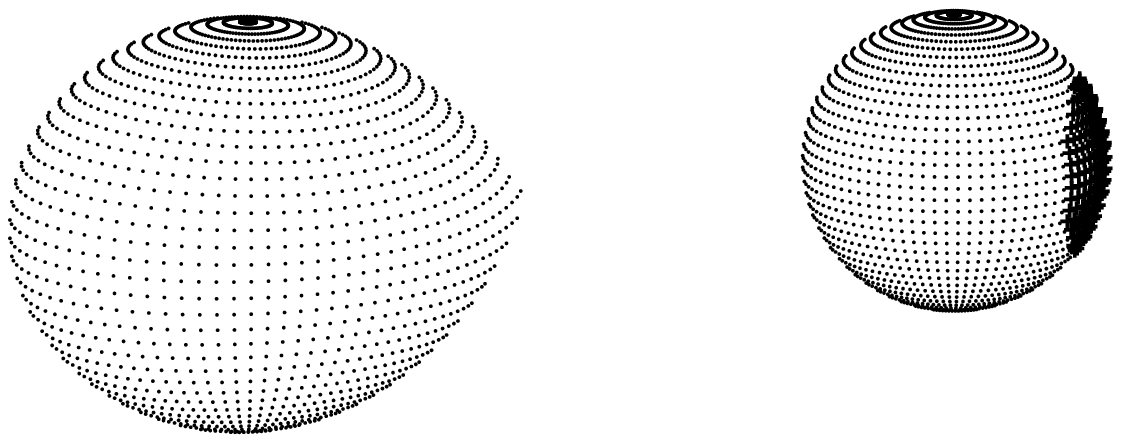}	
	\hspace{0.05in}
	\includegraphics[width=.14\textwidth]{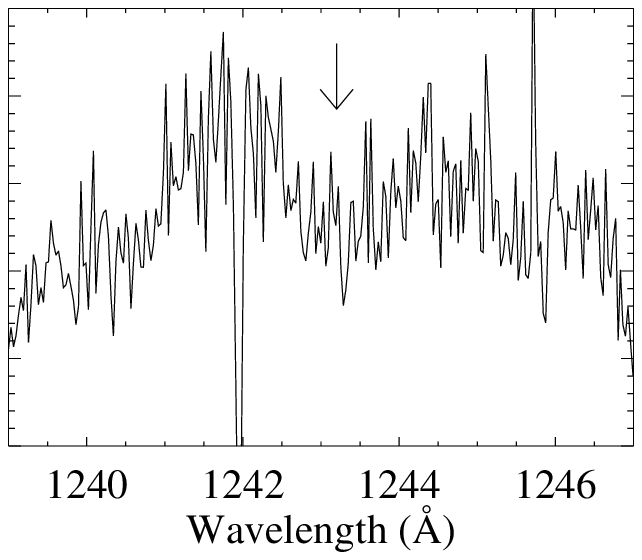}	
	\includegraphics[width=.16\textwidth]{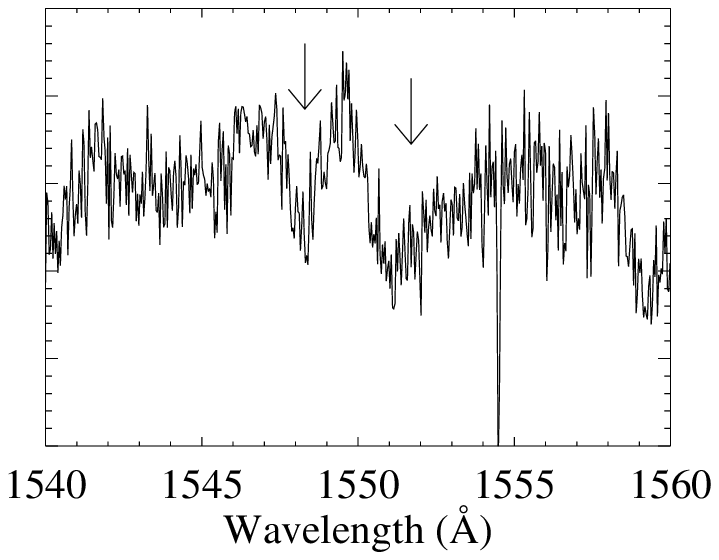}
	\end{array}$
	\end{center}	
	\caption{N \textsc {v} at 1242.80 \AA\ and C \textsc {iv} at 1548.20 \AA\ \& 1550.77 \AA, shown along with the modeled line-of-sight geometry for phases (A) 0.201, (B) 0.395, (C) 0.534, and (D) 0.809, from the 1989 \textit{IUE} data.  The arrows indicate the absorption features in the spectra. We only see N \textsc{v} and C \textsc{iv} absorption when the cool cloud is not eclipsing the primary's hot equator.}
	\label{fig:eclipses-hot}
\end{figure}

The first spectroscopic test of our model is to see if the hot ions are visible only when the primary's equator is not eclipsed by the accretion structure.  Figure \ref{fig:eclipses-hot} shows the profiles of the N \textsc {v} (1242.80 \AA) and C \textsc {iv} (1548.20 \AA) lines at four important phases during the 1989 observations.  The first spectrum (taken at phase 0.201) shows very weak N \textsc {v} and C \textsc {iv} absorption, if any at all, and the corresponding geometry of the model infers that a cool cloud hides the equator of the primary from our view.  The next spectrum, phase 0.395, indicates strengthened N \textsc {v} and C \textsc {iv}, labelled on the graphs with arrows, corresponding to where the accretion cloud dips below our line of sight revealing the primary's hot equator.  Similarly, the hot equator is covered by the cooler accretion cloud near phase 0.534, in the third spectrum of figure \ref{fig:eclipses-hot}, and the hot ions disappear.  The fourth spectrum (phase 0.809) shows very strong N \textsc {v} and C \textsc {iv} absorption,  as we are looking at the region of impact between the gas stream and the star and/or accretion structure.

\begin{figure}
	\includegraphics[width=0.475\textwidth]{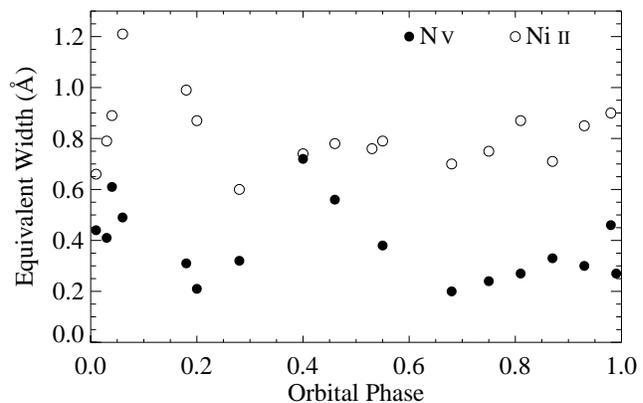}
	\caption{Equivalent widths of the N \textsc{v} (1242.80 \AA) and Ni \textsc{ii} (1454.84 \AA) absorption lines, from the 1989 \textit{IUE} data. The solid circles are for N \textsc{v} and the open circles are for Ni \textsc{ii}.}
	\label{fig:EW-N5-Ni2}
\end{figure}

Another test is to compare the strengths of the hot and cool absorption lines throughout the orbit.  Although the absorption features are badly blended, we did compute equivalent widths (EWs) by direct integration of the normalised data.  Figure \ref{fig:EW-N5-Ni2} displays EW versus orbital phase for the N \textsc{v} and Ni \textsc{ii} absorption profiles, from the 1989 \textit{IUE} data.  Again, the 1989 images cover a full cycle and are ideal for studying orbital variations.  Just after primary minimum, around phase 0.05 to 0.1, the strength of the N \textsc{v} line decreases while that of the Ni \textsc{ii} reaches a maximum.  This is the phase when the accretion structure is most optically thick in our line of sight to the primary.  Later, at around phase 0.4, the N \textsc{v} line strengths reach a maximum while the Ni \textsc{ii} is quite weak.  This, according to our model, is when the hot equator of the primary is exposed to us and the cool accretion structure is mostly below our line of sight.

There is also some spectroscopic evidence for our explanation of R Ara's secular variations.  During times of more direct mass transfer, the impact of the stream may cause a splashing of material over the primary.  This would behave like an asymmetric ``wind'', seen when the splashing material is projected onto the hot star.  Since a stellar wind is detected as a P-Cygni line profile, any asymmetric P-Cygni lines may be caused by this splashing.  The P-Cygni features of 1980, noted by McCluskey and Kondo, appear at phases 0.07 and 0.23 but not at any other phases.  During these phases we are looking at the system from behind the secondary, with a view from above the impact region.  There are no P-Cygni features in the 1989 data even at similar phases.  This is a secular spectroscopic variation supporting our explanation of the secular photometric variations. 

\begin{figure*}
	\begin{center}$
	\begin{array}{ccc}	
	\includegraphics[width=.30\textwidth]{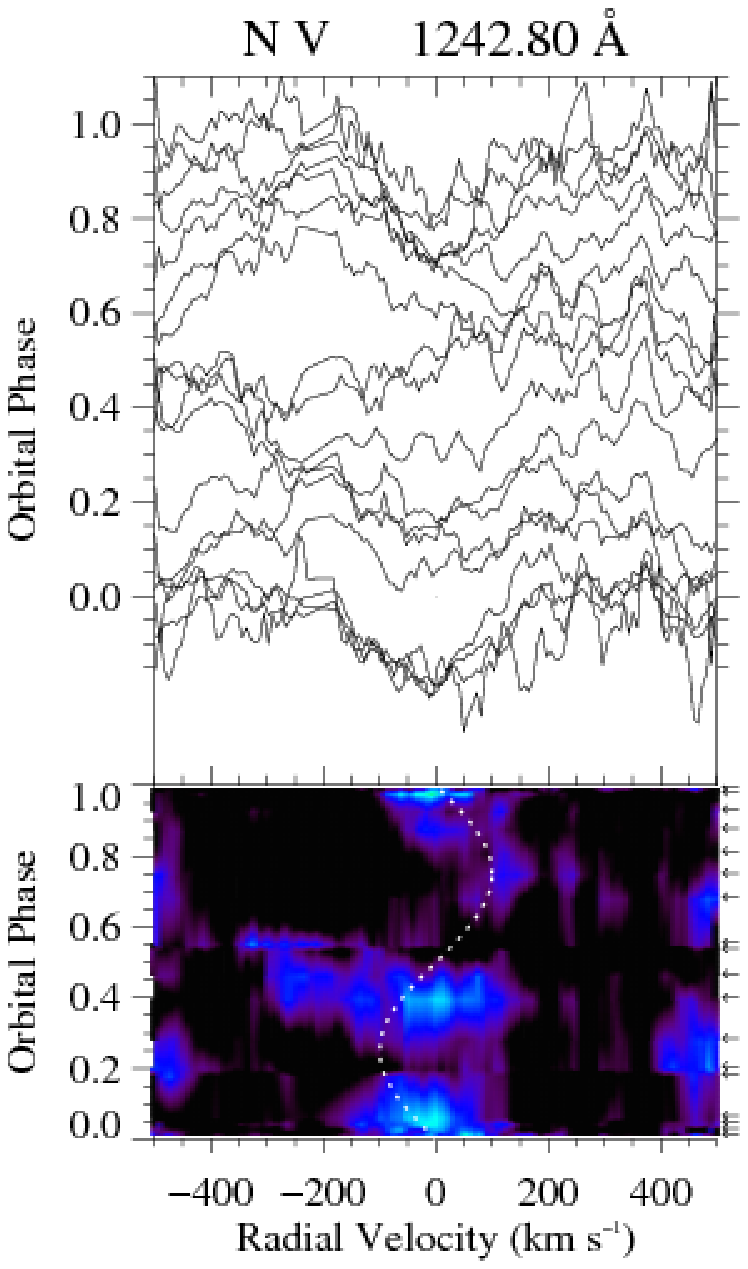}
	\hspace{0.25in}
	\includegraphics[width=.30\textwidth]{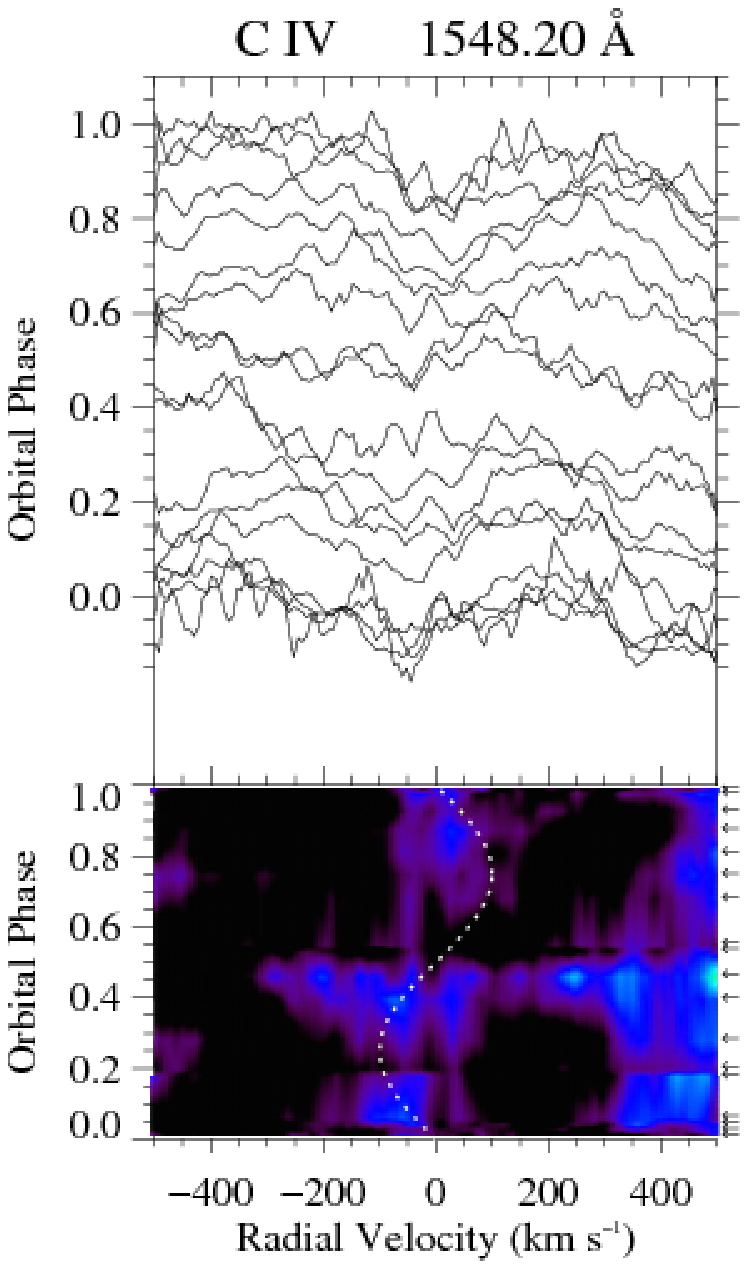}
	\hspace{0.25in}
	\includegraphics[width=.30\textwidth]{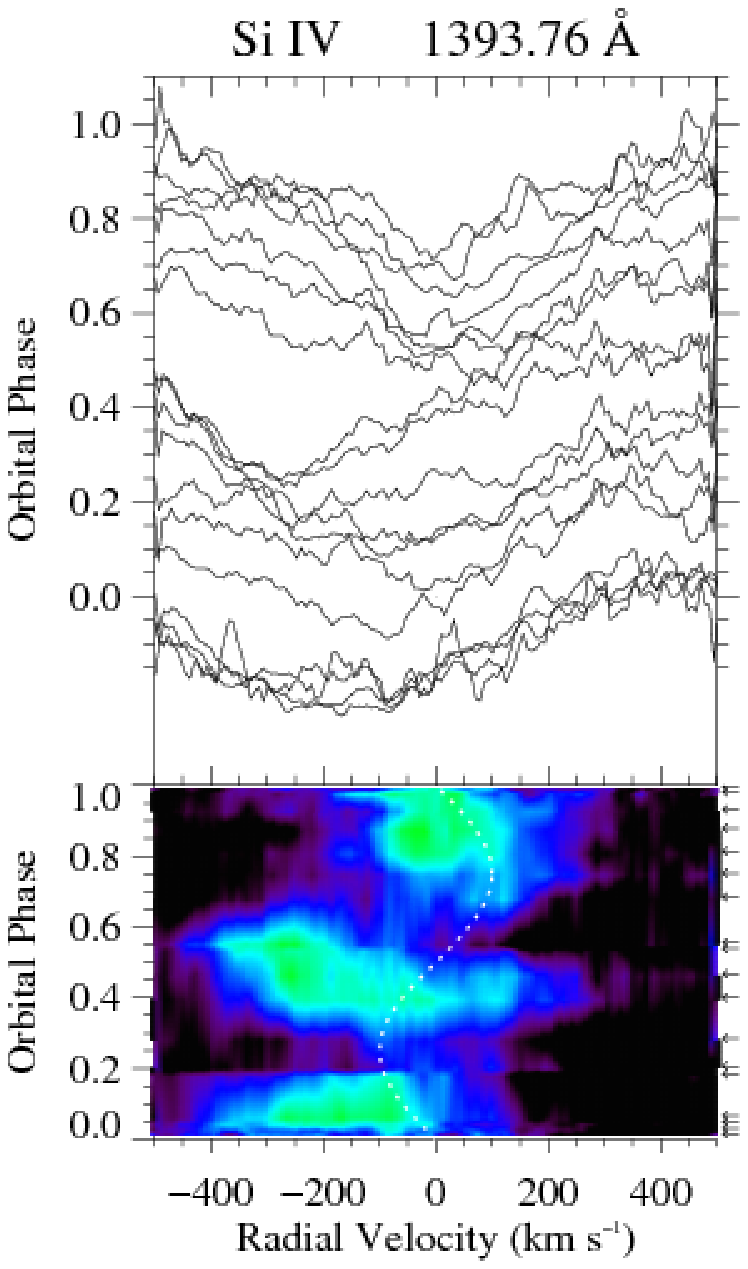}
	\end{array}$
	\end{center}	
	\caption{These are colour scale RV plots for N \textsc{v}, C \textsc{iv}, and Si \textsc{iv} in 1989.  These images show the motions of the hotter regions of R Ara.  The arrows indicate actual data and the white dotted curve represents the estimated 100 $\frac{km}{s}$ orbital motion of the centre of mass of the primary star.}
	\label{fig:N5-C4-gray}
\end{figure*}

\begin{figure*}
	\begin{center}$
	\begin{array}{ccc}	
	\includegraphics[width=.30\textwidth]{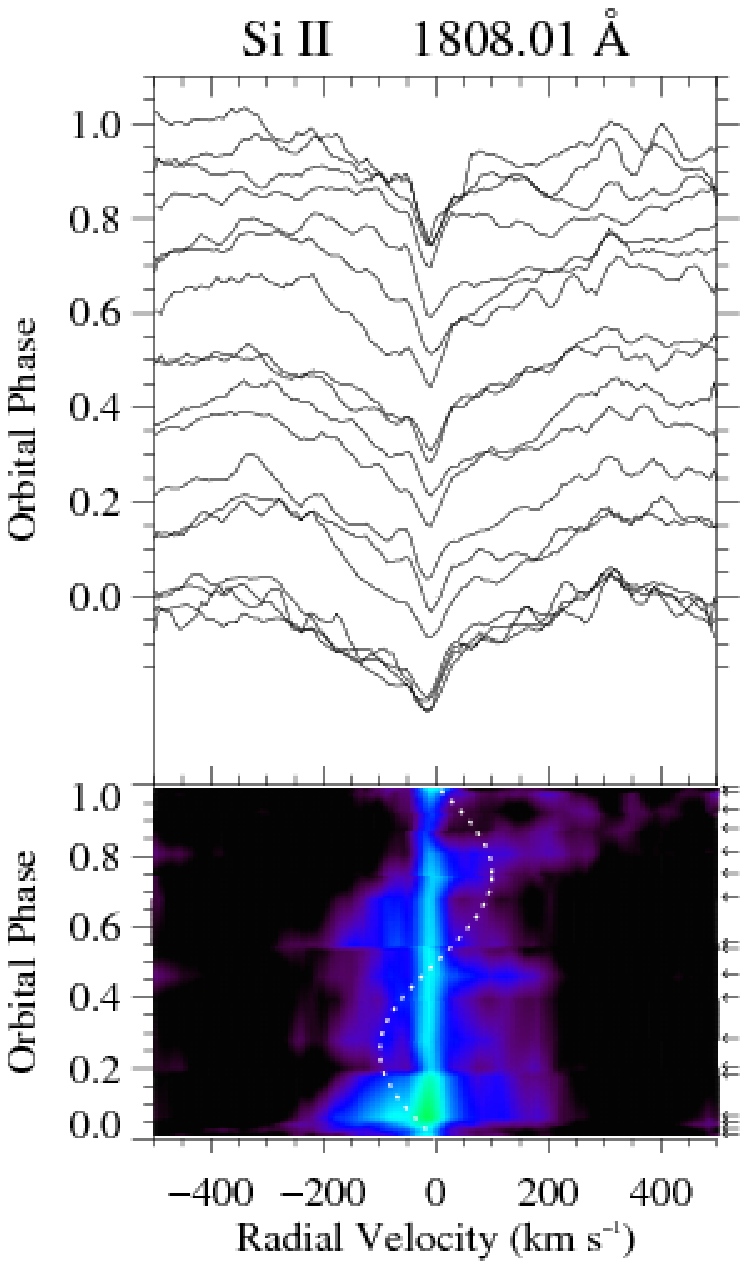}	
	\hspace{0.25in}
	\includegraphics[width=.30\textwidth]{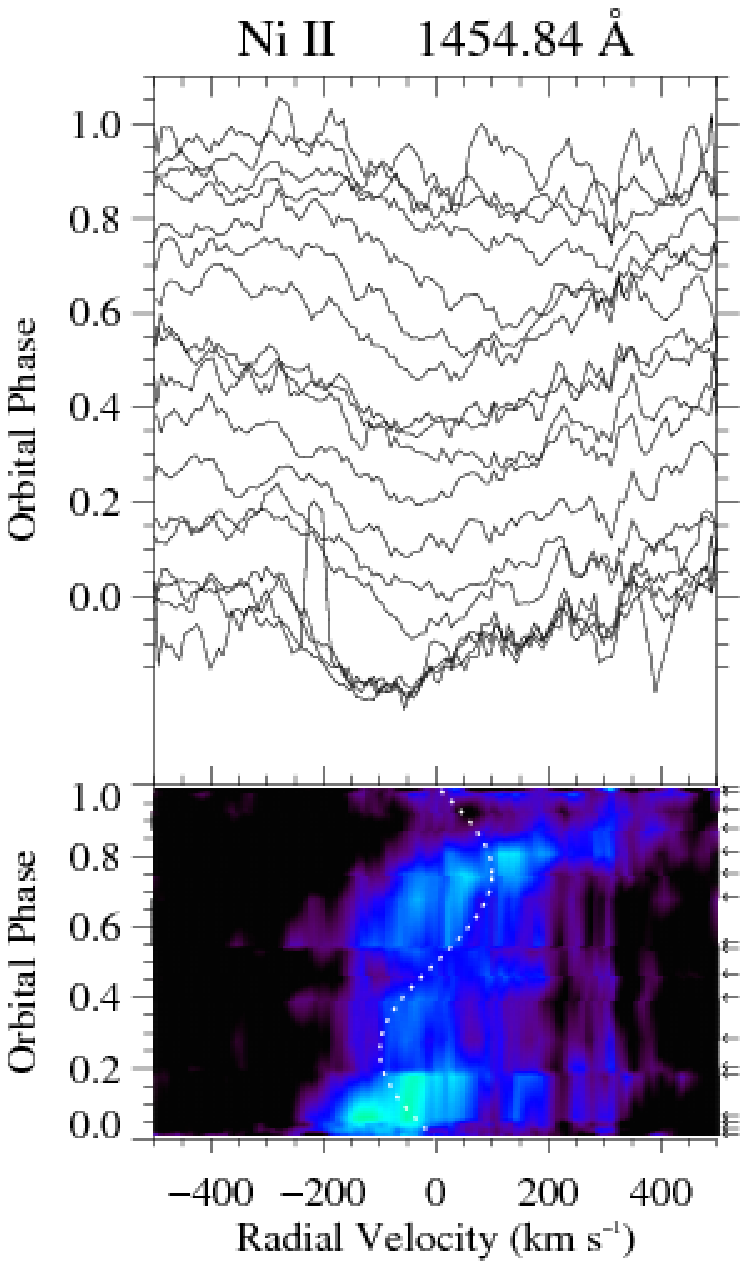}
	\hspace{0.25in}
	\includegraphics[width=.30\textwidth]{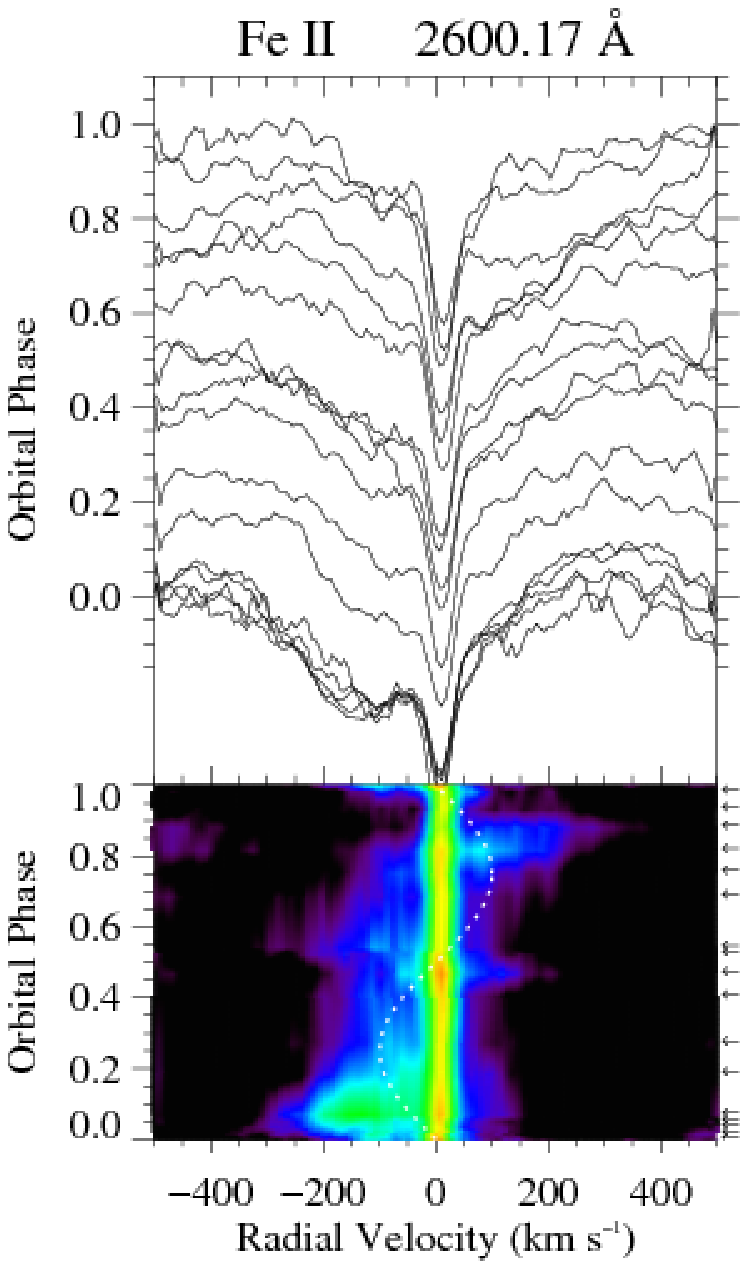}
	\end{array}$
	\end{center}	
	\caption{These are colour scale RV plots for Si \textsc{ii}, Ni \textsc{ii}, and Fe \textsc{ii} in 1989.  These images show the motions of the cooler regions of R Ara.  The arrows indicate actual data and the white dotted curve represents the estimated 100 $\frac{km}{s}$ orbital motion of the centre of mass of the primary star.}
	\label{fig:Si2-Ni2-Fe2-gray}
\end{figure*}

An RV analysis provides the most rigorous test of our photometric model by answering the question asked at the end of section \ref{Sec:UV Light Curve Analysis}.  Due to the complicated blending of the absorption lines in R Ara's spectrum, we employed colour scale RV plots to visualize the dynamic behavior of the ions within the system.  Such plots are displayed in figure \ref{fig:N5-C4-gray}, for the hotter ions, and figure \ref{fig:Si2-Ni2-Fe2-gray}, for the cooler ions.  The 1989 data was used due to the good phase coverage over a single orbit.  The arrows to the right of the colour scale RV plots indicate actual data points with a linear interpolation of the data between them.  The dotted white line in these plots is not a fit to the data, but it is the estimated motion of the centre of mass of the primary star ($k\ =\ 100\ km\ s^{-1}$) and it is plotted to aid in distinguishing orbital from non-orbital motion.


The ``hottest'' ion, N \textsc {v}, clearly follows the orbital motion of the primary star during the last quarter of the orbit, with the obvious deviation of the large shortward shifted ``flare'' near phase 0.4.  Si \textsc{iv} also shows a clear orbital component during the last quarter of the orbit, and it exhibits the flare before the secondary eclipse.  This flare will be discussed later with the 1985 data, as they are concentrated around this portion of the orbit and provide better time resolution allowing us to ``zoom-in'' to that feature.

The ``cooler'' ions exhibit a sort of ``S-shape'' on the colour scale RV plots, which is characteristic of an accretion disc. They show high receding velocities before primary eclipse followed by high approaching velocities after the eclipse.  Also, they lack the mid-orbit flare that the hotter ions present.  Si \textsc{ii} and Fe \textsc{ii} are cool enough to be found in the interstellar medium (ISM), and these colour scale RV plots clearly show their interstellar components as vertical lines.

The differences between the motions of the hot and cool regions within R Ara yield strong support for our model.  We expect the cooler ions to follow the motion of a disc or cloud moving around the primary.  We also expect the hottest ions to be present at the primary's equator, as the equitorial hot river is driven by the impact of the mass-transfer stream.

\begin{figure*}
	\begin{center}$
	\begin{array}{cccc}	
	\multirow{4}{*}{\includegraphics[width=.30\textwidth]{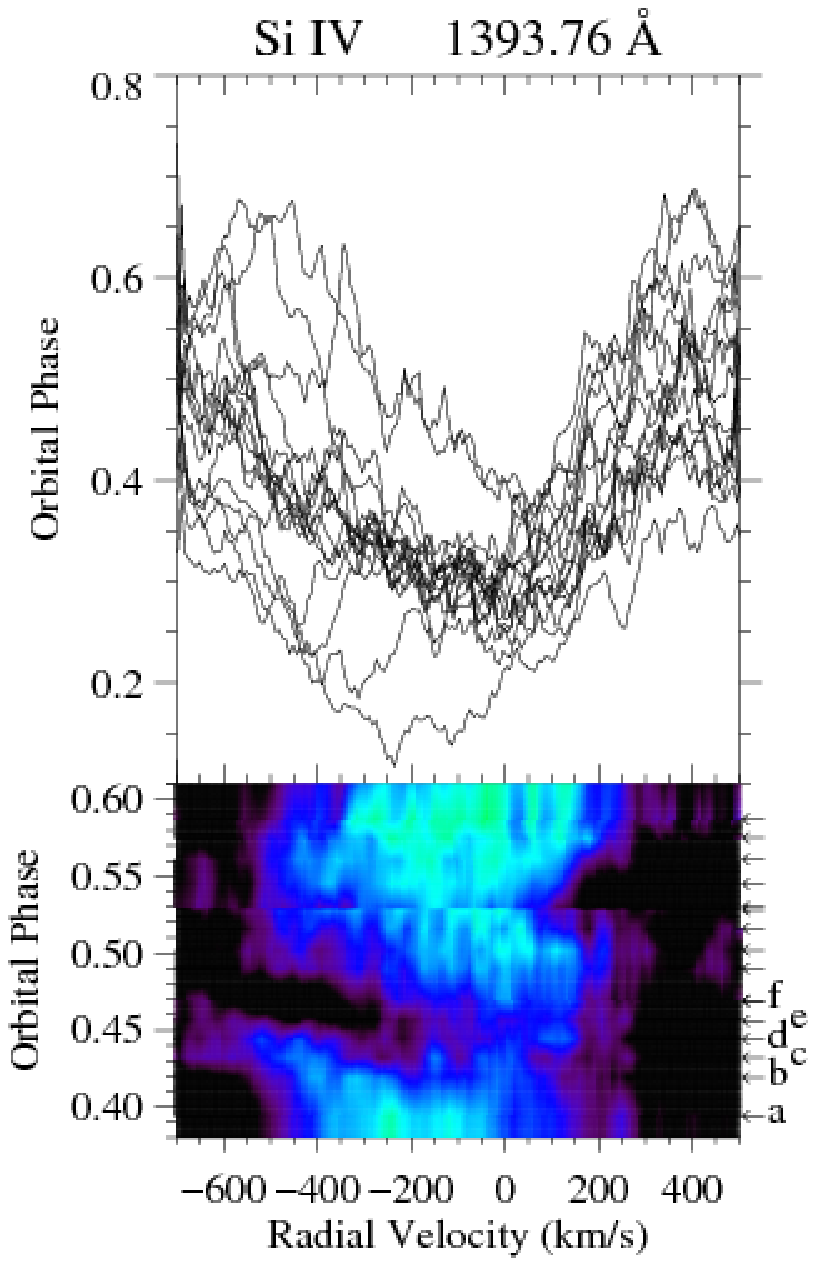}}
	&\includegraphics[width=.14\textwidth]{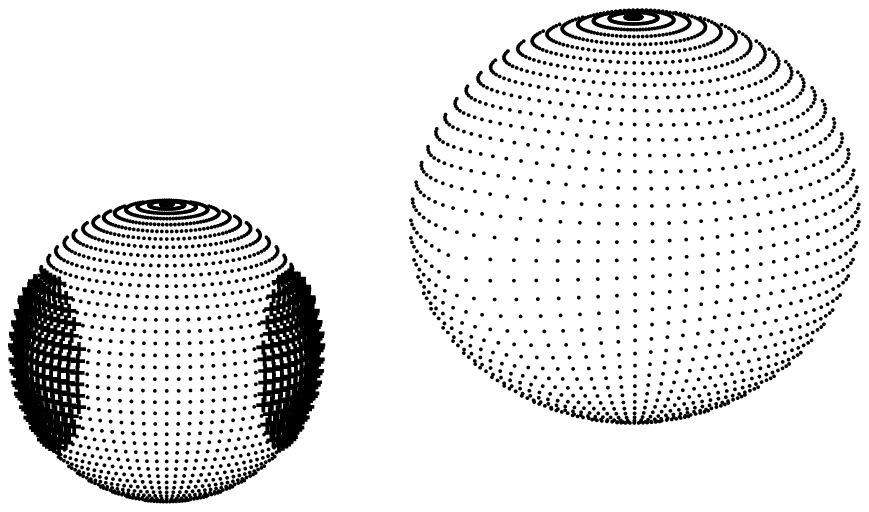}
	\hspace{0.3in}	
	\includegraphics[width=.13\textwidth]{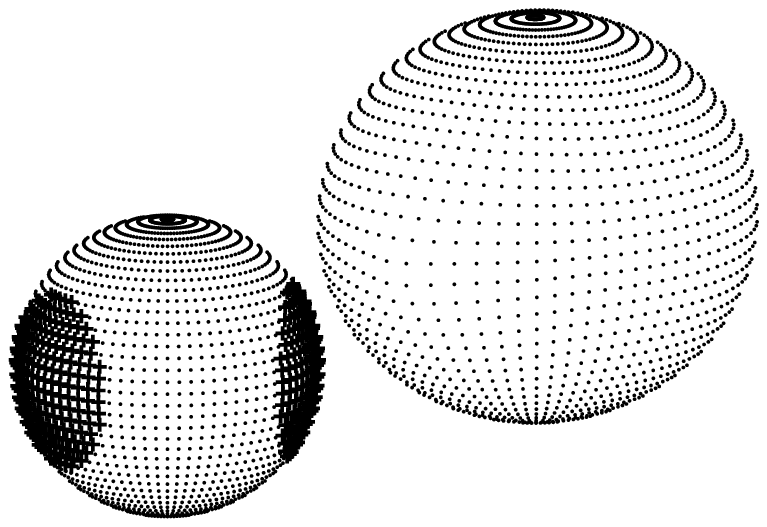}
	\hspace{0.3in}
	\includegraphics[width=.12\textwidth]{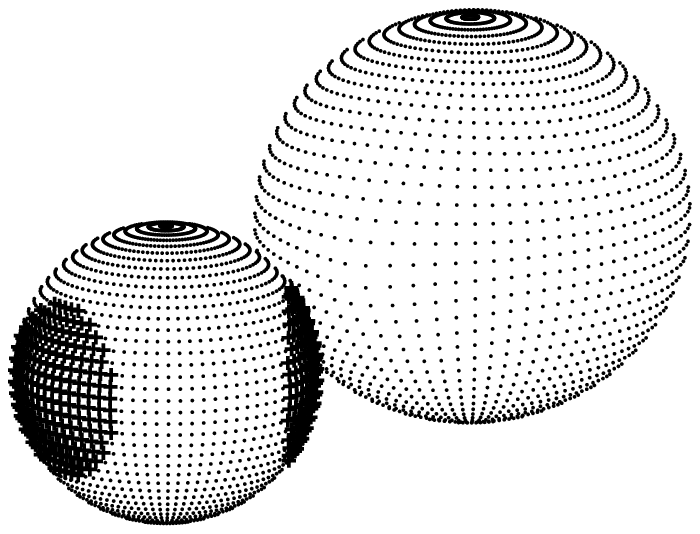}\\
	\vspace{0.2in}
	&\includegraphics[width=.19\textwidth]{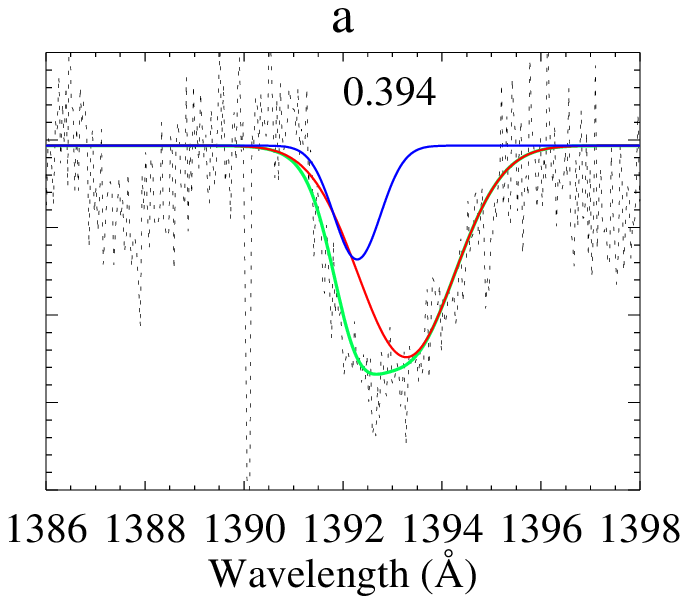}	
	\includegraphics[width=.19\textwidth]{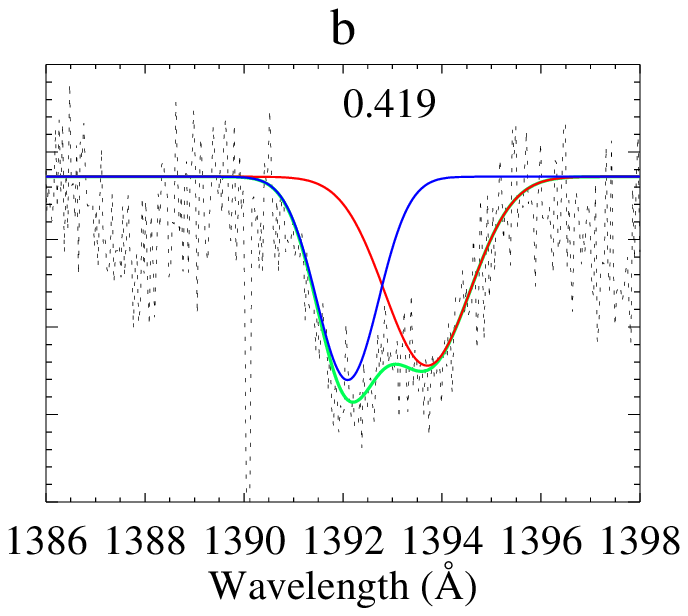}	
	\includegraphics[width=.19\textwidth]{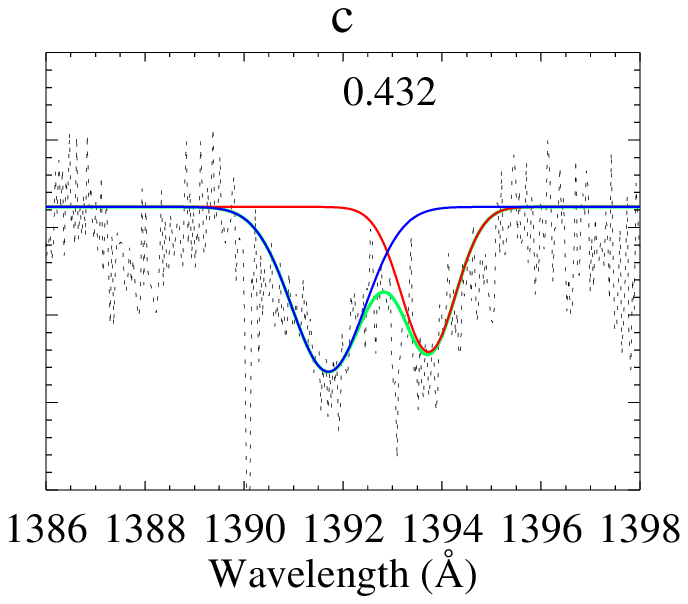}\\
	&\includegraphics[width=.11\textwidth]{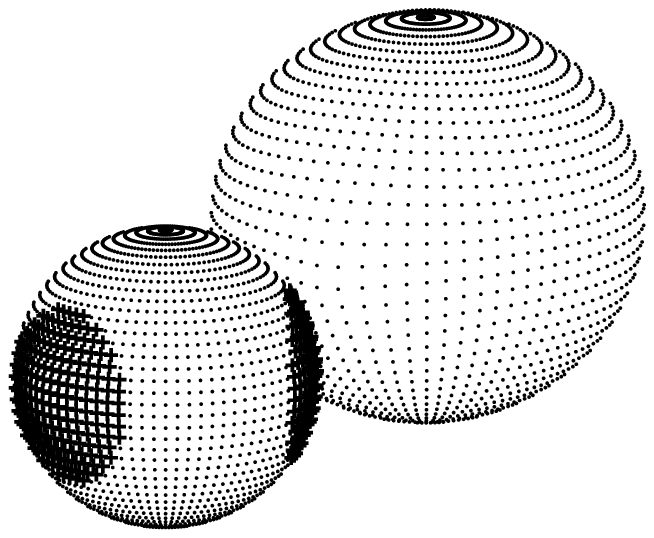}
	\hspace{0.5in}	
	\includegraphics[width=.10\textwidth]{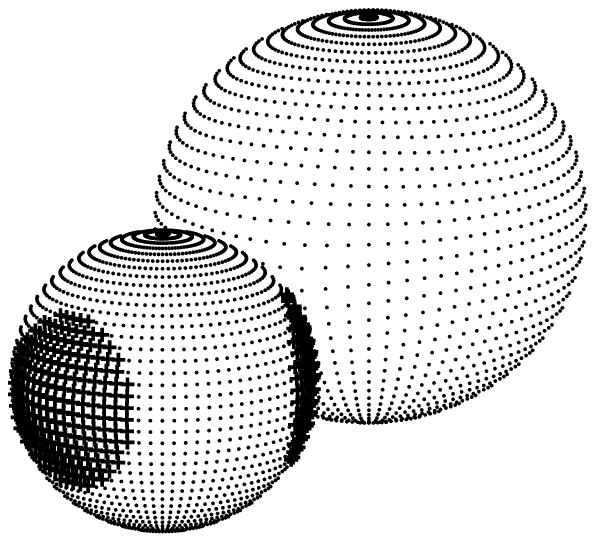}	
	\hspace{0.5in}
	\includegraphics[width=.09\textwidth]{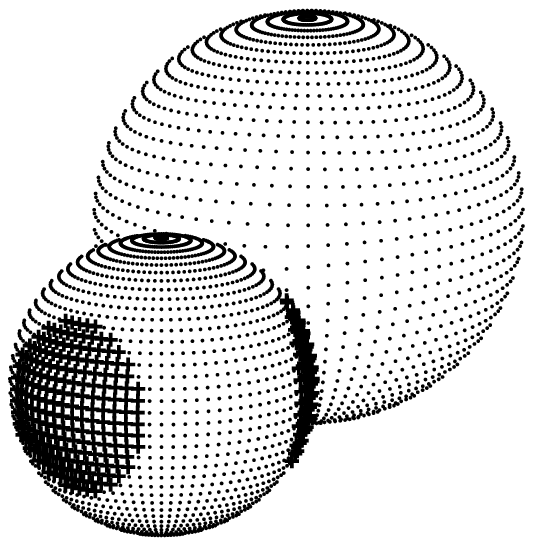}\\
	&\includegraphics[width=.19\textwidth]{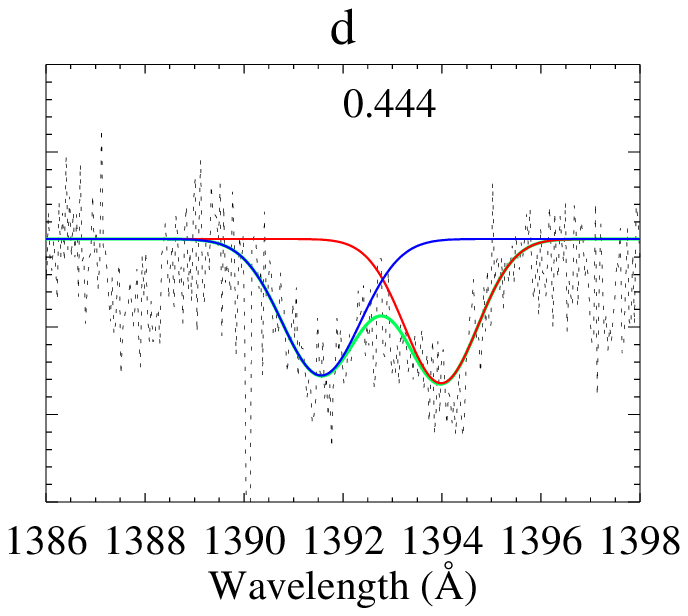}	
	\includegraphics[width=.19\textwidth]{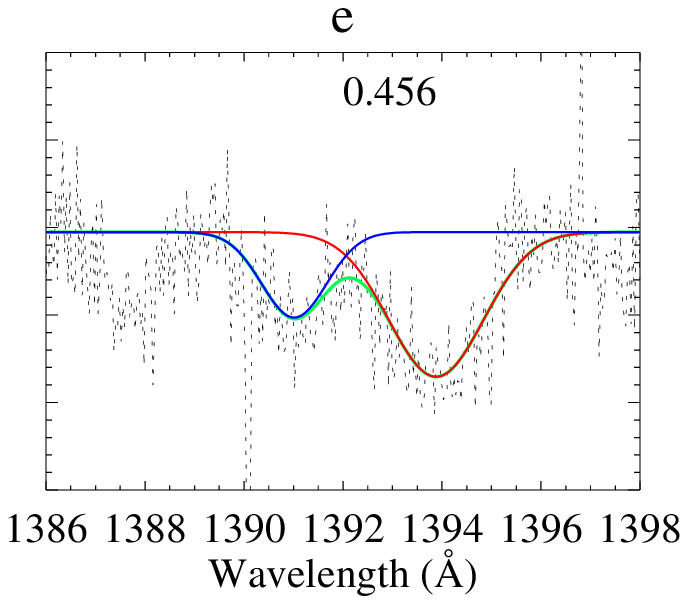}	
	\includegraphics[width=.19\textwidth]{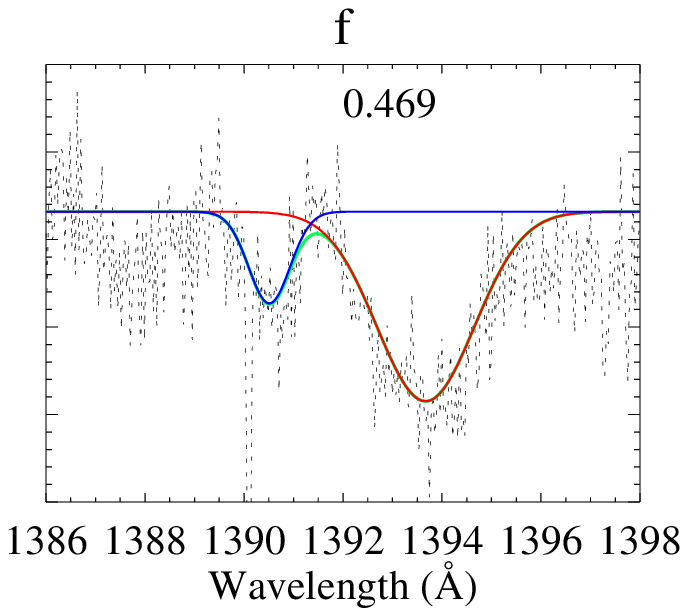}
	\end{array}$
	\end{center}	
	\caption{The ``Splitting'' of Si \textsc{iv} absorption line profiles in 1985.  The colour scale RV plot on the left is similar to those in figures \ref{fig:N5-C4-gray} and \ref{fig:Si2-Ni2-Fe2-gray}, but focused between phases 0.4 and 0.6.  The time resolution of the 1985 observations is very good, allowing us to track the two components as blended Gaussians.  The component corresponding to Si \textsc{iv} in the accretion structure is diplayed in red and the Si \textsc{iv} located at the primary's hot equator is in blue.  The two components blend to give the green profile which matches the data.  The line-of-sight geometry from the BM3 model is given for each image, and the phase value is printed above each line profile.}
	\label{fig:85splitting}
\end{figure*}

We will now zoom in to the flare seen in the RV plots for the hotter ions.  Specifically, we will consider the Si \textsc{iv} line profiles of 1985, which actually display a clear splitting of the line into two blended components.  Si \textsc{iv} always gives us the strongest, sharpest lines, probably due to its high abundance.  In figure \ref{fig:85splitting}, we diplay the colour scale RV plot for Si \textsc{iv} in 1985, focused on phases 0.4 to 0.6.  Here we see the flare in much more detail.  The observations labelled \textit{a} through \textit{f} in the figure are modeled with two Gaussian-shaped absorption lines blended together.  The figure also shows the geometry of the system at the time of each image, including the ``cool clouds'' from the BM3 model.

\begin{figure}
	\includegraphics[width=.475\textwidth]{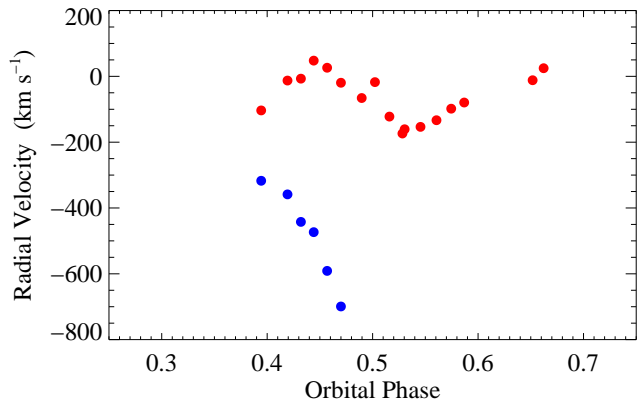}
	\caption{The radial velocities of the two components of the blended Si \textsc{iv} line in figure \ref{fig:85splitting}.  The blue points are the RVs of the blue Gaussian and the red points are for the red component.}
	\label{fig:85BlendRVs}
\end{figure}

According to our model, we are looking at both the accretion structure (which is descending below our line of sight to the centre of the primary) and the primary's equator (in the centre of our line of sight) during phases \textit{a} to \textit{f} of figure \ref{fig:85splitting}.  Figure \ref{fig:85BlendRVs} tracks the RVs of the two components through the 1985 observations.  The blue points in the plot represent the Gaussian component modeled in blue in figure \ref{fig:85splitting} and the red points are for the red Gaussian.
The red component shows RVs that vary as we would expect the accretion disc to vary and is likely arising from (the warmer parts of) the accretion structure, which appears to oscillate between receding and approaching.  The blue Gaussian in the figure moves steadily to higher and higher approaching velocities in the progression from \textit{a} to \textit{f}, and it is most likely arising from the star's hot equator as we are looking more and more ``down the barrel'' of the gas stream at each of those phases.  In order the have a hot region of the primary's photosphere reach velocities of 800 $km\ s^{-1}$, there must be a mass transfer stream driving it.

\begin{figure*}
	\begin{center}$
	\begin{array}{cc}	
	\includegraphics[width=.35\textwidth]{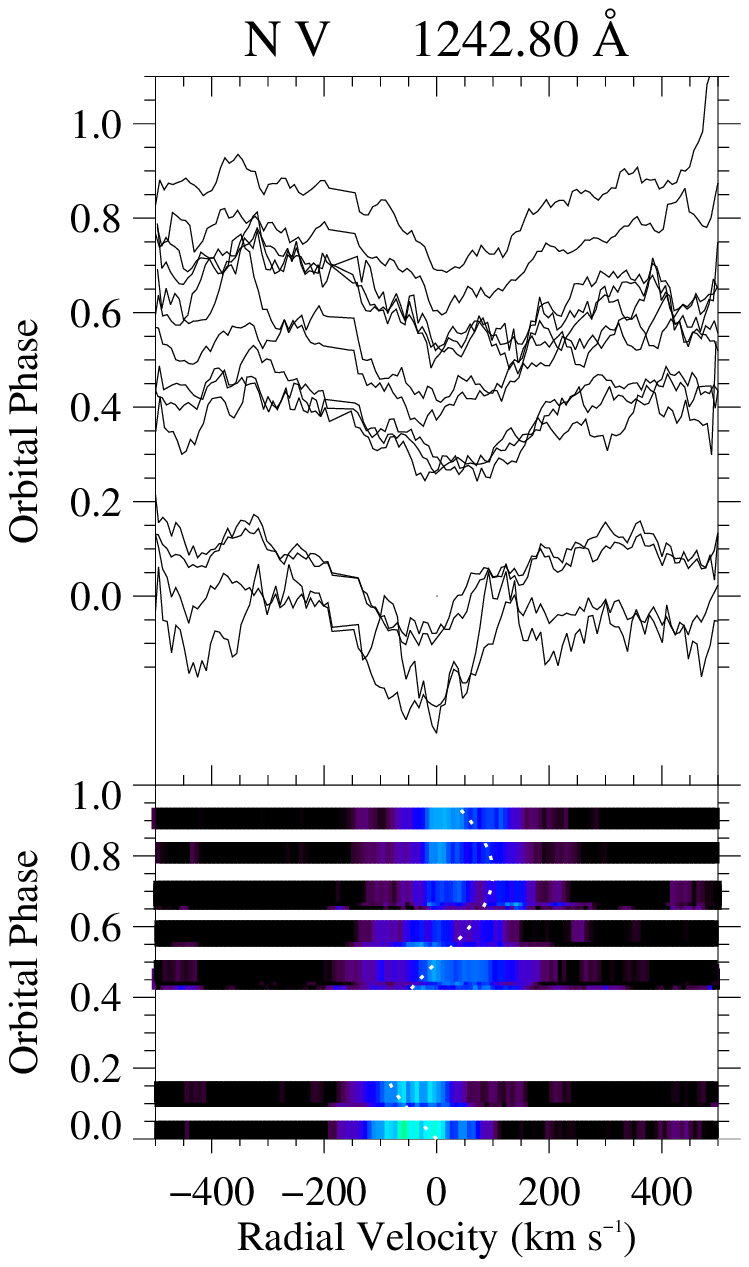}	
	\hspace{0.25in}
	\includegraphics[width=.35\textwidth]{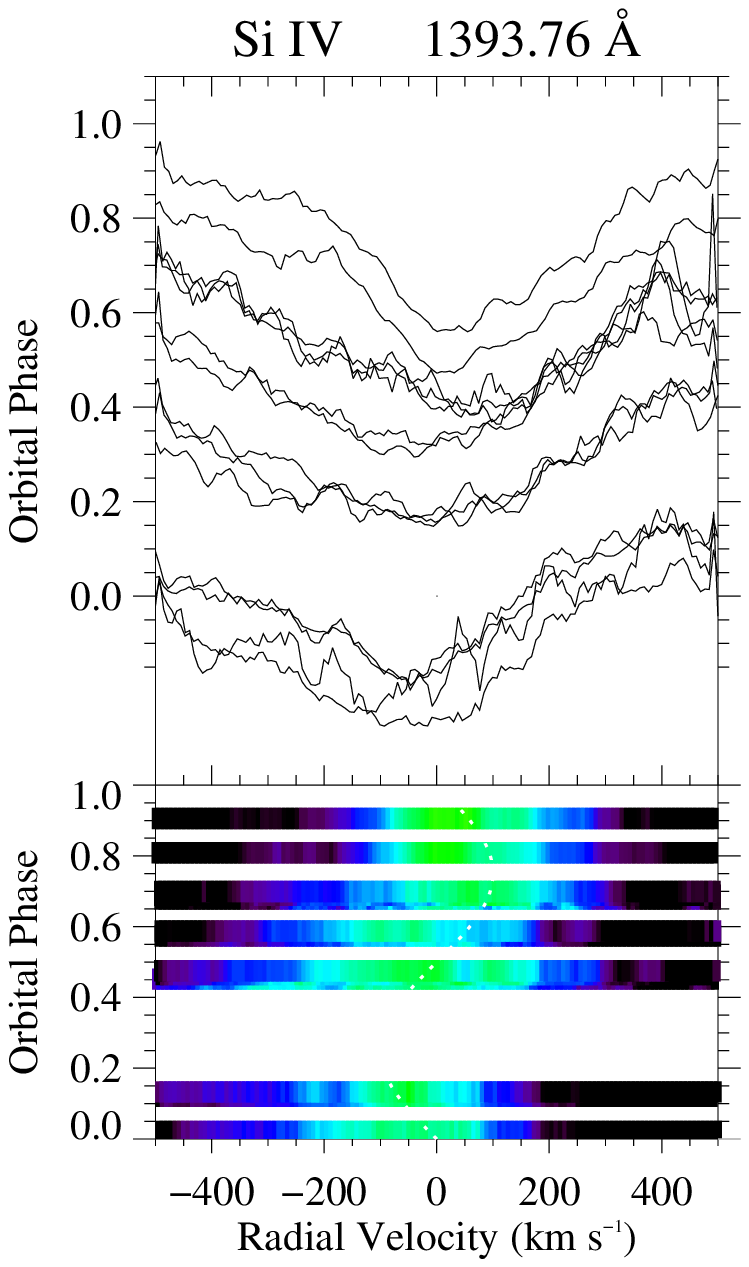}\\
	\includegraphics[width=.35\textwidth]{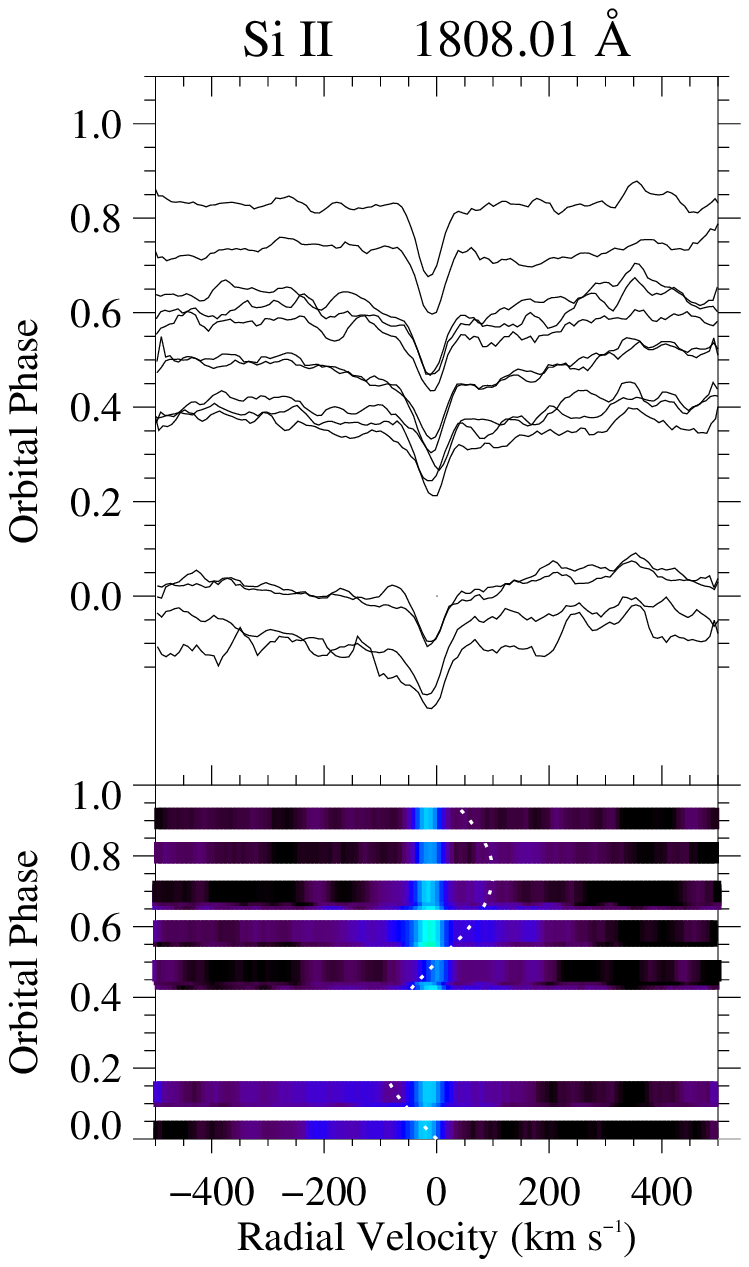}
	\hspace{0.25in}
	\includegraphics[width=.35\textwidth]{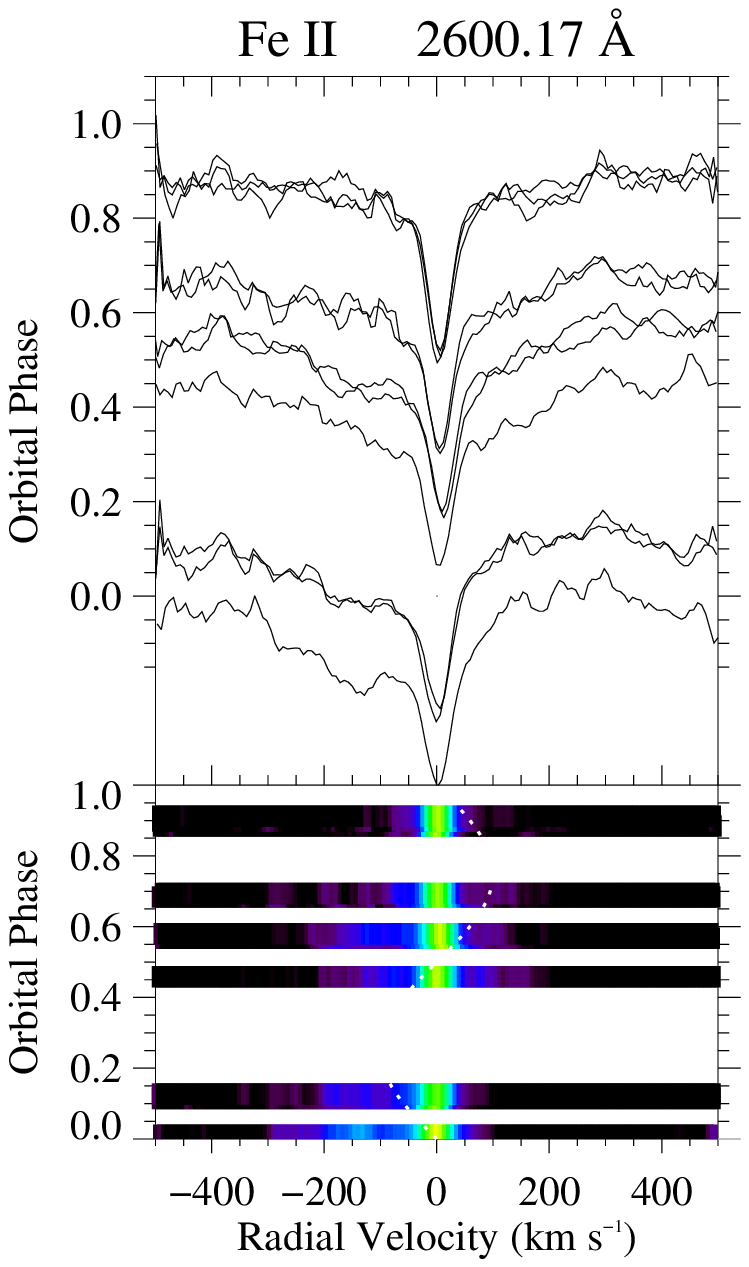}
	\end{array}$
	\end{center}	
	\caption{These are colour scale RV plots for N \textsc{v}, Si \textsc{iv}, Si \textsc{ii}, and Fe \textsc{ii} in 1982.  These data are not dense enough to be interpolated, so they apear as individual strips for each image.  The white dotted curve represents the estimated 100 $\frac{km}{s}$ orbital motion of the centre of mass of the primary star.}
	\label{fig:All-82-color}
\end{figure*}

Our final spectroscopic test of the photometric model is to compare the absorption line strengths and RVs in 1982 with those in 1989.  Figure \ref{fig:All-82-color} displays colour scale RV plots for N \textsc{v} and Si \textsc{iv}, the hotter ions, as well as for Si \textsc{ii} and Fe \textsc{ii}, the cooler ones, in 1982.  Since the 1982 observations are fewer in number and not as frequent as in 1989, these plots are not interpolated, rather they consist of individual strips for each image.  The first thing we notice is that the cooler ions are much weaker in 1982 than in 1989.  In fact, mostly all of the Si \textsc{ii} and Fe \textsc{ii} absorption is by the ISM in 1982.  This is further proof for our explanation of the photometric secular differences between 1982 and 1989.  Recall that the system is brighter in 1982 than in 1989, resulting from more direct mass transfer and a thinner accretion structure.
In addition to seeing less of the cool accretion structure, we would expect to see more of the primary's hot equator in 1982.  Indeed, the absorption by the hotter ions is stronger in 1982, and those ions are present for more of the orbit.  We can even see a component of the N \textsc{v} line following the estimated orbital motion of the primary during every observed phase in 1982.  In 1982, we are certainly seeing more of the hot equator of the primary's photosphere and less of the cool accretion structure than in 1989.

\section{Discussion / Evolutionary Status}
\label{sec:Descussion}

Interacting binaries consisting of a cool (F-K) secondary donating mass to a hotter (B-A) and less evolved primary by RLOF are considered to be Algols.  But, as is pointed out by \citet{Albright1996}, Algols are in a slow and steady phase of mass transfer ($\dot{M} \sim 10^{-11}$ to $10^{-7}\ M_{\odot}\ yr^{-1}$), which is why they are used as laboratories for studying accretion processes in binary stars.  Is R Ara an Algol?  If not, in what stage of its evolution does it currently reside?

We have explained R Ara's photometric variations and its complicated spectrum as the result of unsteady accretion via a variable accretion structure.  We can estimate the mass transfer rate using the new 2008 PROMPT observations.  We will use Nield's ephemeris to calculate the period change rate and mass transfer rate between 1986 and 2008.  The calculated time of primary minimum is C = HJD 2,454,541.547 and the observed time is O = HJD 2,454,541,757.  The number of orbits between Nield's ephemeris and the 2008 eclipse is $n = 1798$ and the period from the Nield ephemeris is P = 4.425132 days.  The rate of period increase between the two epochs is then:
	\begin{eqnarray}
	\dot{P} &=& \frac{2(O-C)}{n^{2}P} \\
	&=& 2.94 \times 10^{-8}\ (\pm\ 2.80 \times 10^{-10})\ \frac{days}{day}
	\end{eqnarray}
From this we find $\dot{P}/P = 2.43 \times 10^{-6}\ yr^{-1}$, which corresponds to a timescale of roughly $4 \times 10^{5}$ years for this phase of R Ara's evolution; a very short-lived stage.

If we assume the masses of the stars to be $M_{1} = 4 M_{\odot}$ and $M_{2} = 1.4 M_{\odot}$ (these are Sahade's original estimates, but they are also consistent with the mass ratios used both in Nield's photometric model and our photometric model in this paper), and if we assume conservative mass transfer, we can use the equation from \citet{Kwee1958} to get a mass transfer rate of:
	\begin{eqnarray}
	\dot{M} &=& \frac{\dot{P}M_{1}M_{2}}{3P(M_{1}-M_{2})} \\
	&=& 1.74 \times 10^{-6}\ (\pm\ 1.66 \times 10^{-8})\  \frac{M_{\odot}}{yr} 
	\end{eqnarray}
which is much more rapid than typical Algols.  In fact, at this rate of mass transfer the secondary would lose half of its current mass in just a few hundred thousand years.  

We believe R Ara is in the pre-Algol stage of its evolution, just past the reversal of mass ratio, which would be characterised by rapid mass transfer and an unstable, transient accretion structure.  We project that R Ara's mass transfer rate will eventually slow down as it enters a more stable state.  At that time, R Ara may become a classic Algol type binary star.

Many frequent times of primary minima are desired in order to more accurately track a change in the orbital period of R Ara and to ultimately conduct a complete O-C analysis.  Spectroscopic data for use in constructing Doppler tomograms would also be of great benefit to visualize R Ara's mass flow and accretion structure.

\end{document}